\documentclass[11pt]{article}

\usepackage{epsfig,amssymb}
\newcommand{\be}{\begin{equation}}
\newcommand{\ee}{\end{equation}}

\makeatletter
\def\eqalign#1{\null\vcenter{\def\\{\cr}\openup\jot\m@th
  \ialign{\strut$\displaystyle{##}$\hfil&$\displaystyle{{}##}$\hfil
      \crcr#1\crcr}}\,}
\makeatother

\newcommand{\tr}{\mathrm{tr}\,}
\newcommand{\la}{\label}

\newcommand{\wt}{\widetilde}
\newcommand{\de}{\delta}

\newcommand{\al}{\alpha}
\newcommand{\bt}{\beta}
\newcommand{\ga}{\gamma}

\newcommand{\si}{\sigma}
\newcommand{\Si}{\Sigma}
\newcommand{\om}{\omega}
\newcommand{\Om}{\Omega}
\newcommand{\lb}{\lambda}
\newcommand{\ze}{\zeta}

\renewcommand{\th}{\theta}

\newcommand{\ep}{\varepsilon }

\def\bbc{\mathbb C}
\def\bbr{\mathbb R}

\begin{document}
\bigskip\bigskip\bigskip
\begin{center}
{\Large\bf
Asymptotics for a determinant with a confluent hypergeometric kernel
}\\
\bigskip\bigskip\bigskip
\centerline{ P. Deift\footnote{Courant Institute of Mathematical Sciences, New York, NY 10003, USA}, 
I. Krasovsky$^2$, and J. Vasilevska\footnote{
 Department of Mathematical Sciences, Brunel University, Uxbridge UB83PH, United Kingdom}}
\end{center}
\bigskip\bigskip\bigskip
\noindent{\bf Abstract.}
We obtain ``large gap'' asymptotics for a Fredholm determinant with
a confluent hypergeometric kernel. We also obtain asymptotics for 
determinants with two types of Bessel kernels which appeared in random 
matrix theory. 


\section{Introduction}
Let $K^{(\al,\bt)}$ be the operator acting on $L^2(-s,s)$, $s>0$, with kernel
\be
\label{kernel}
K^{(\alpha,\beta)}(u,v)=\frac 1{2\pi i}\frac{\Gamma(1+\alpha+\beta)
\Gamma(1+\alpha-\beta)}{\Gamma(1+2\alpha)^2}
\frac{A(u)B(v)-A(v)B(u)}{u-v},
\ee
where
\begin{eqnarray}
&A(x)=g_\beta^{1/2}(x)|2x|^{\alpha}e^{-ix}\phi(1+\alpha+\beta,1+2\alpha,2ix),&\nonumber\\
&B(x)=g_\beta^{1/2}(x)|2x|^{\alpha}e^{ix}\phi(1+\alpha-\beta,1+2\alpha,-2ix),&\nonumber\\
&g_\beta(x)=\cases{
e^{-\pi i\beta},& $x\ge 0$, \cr
e^{\pi i\beta}, & $x<0$. 
},\qquad \alpha,\beta\in\bbc,\quad \Re{\alpha}>-1/2,\quad \al\pm\bt\neq-1,-2,
\dots&\nonumber
\end{eqnarray}
Here $\Gamma(x)$ is Euler's $\Gamma$-function, and $\phi(a,c,z)$ is the confluent hypergeometric 
function (see, e.g., \cite{Abr})
\begin{equation}\label{phidef}
\phi(a,c,z) = 1+\sum_{n=1}^{\infty}\frac{a(a+1)\cdots(a+n-1)}{c(c+1)\cdots(c+n-1)}\frac{z^n}{n!}.
\end{equation}
Using the standard recurrence formulae for $\phi(a,c,z)$ (see (\ref{recurrence}) below),
we can rewrite (\ref{kernel}) in another form:\footnote{
The case $\al=0$ is understood here as a limit $\al\to 0$.}
\be\la{kernel2}
\eqalign{
K^{(\al,\bt)}(u,v)=\frac{1}{\pi}\frac{\Gamma(1+\al+\bt)\Gamma(1+\al-\bt)}
{(1+2\al)\Gamma(1+2\al)^2}
g^{1/2}_\bt(u)g^{1/2}_\bt(v)e^{-i(u+v)}\frac{4^{\al}|uv|^\al}{u-v}\\
\times [u\phi(1+\al+\bt,2+2\al,2iu)\phi(\al+\bt,2\al,2iv)
-v\phi(1+\al+\bt,2+2\al,2iv)\phi(\al+\bt,2\al,2iu)].}
\ee

The kernel (\ref{kernel}) or (\ref{kernel2}) is called the confluent hypergeometric kernel.
For $\al\in\bbr$, $\bt\in i\bbr$ (in this case the kernel is real,
which is easy to see from (\ref{kernel})), 
it was considered by Borodin and Olshanski in \cite{BO}, 
and by Borodin and Deift \cite{BD} (Proposition 8.13). 
This kernel arises in several different, but related, contexts:

First, following \cite{BO},
consider the space $H$ of infinite Hermitian matrices $(H_{jk})_{j,k=1}^{\infty}$. The $U(\infty)$, 
the inductive limit of the unitary groups $U(N)$, $N\to\infty$, acts on $H$ by conjugations. 
A probability Borel measure on $H$ which is invariant under the action 
of $U(\infty)$ is called ergodic, if any invariant mod 0 set has measure 0 or 1. 
Consider the space $\Om$ whose elements consist of 2 infinite sequences,
\[
\al_1^+\ge\al_2^+\ge\cdots\ge 0,\qquad  \al_1^-\ge\al_2^-\ge\cdots\ge 0,
\qquad
\mbox{where}\qquad
\sum_{j=1}^{\infty}(\al_j^+)^2+\sum_{j=1}^{\infty}(\al_j^-)^2<\infty,
\]
together with 2 extra real parameters $\ga_1$, $\ga_2$, where $\ga_2\ge 0$.
It turns out that the elements of $\Om$ parametrize the ergodic measures on  $H$. Furthermore, it can 
be proved that any $U(\infty)$-invariant probability measure on $H$ decomposes on ergodic components, 
i.e., it can be written as a continuous convex combination of ergodic measures. This {\it spectral decomposition}
is unique and is determined by a probability measure on $\Om$ which is called the {\it spectral measure}
of the original invariant measure.  

The space $\Om$ maps to the space $\mathrm{Conf}(\bbr^*)$ of point configurations on the punctured 
real line $\bbr^*=\bbr\setminus\{ 0 \}$ in the following way:
\[
(\{\al_j^+\}_{j=1,2,\dots},\{\al_j^-\}_{j=1,2,\dots},\ga_1,\ga_2)\to 
(-\al_1^-,-\al_2^-,\dots,\al_2^+,\al_1^+),
\]
where possible zeros among $\al_j^{\pm}$ are omitted. Under this map spectral measures corresponding 
to invariant measures on $H$, push-forwards to measures on $\mathrm{Conf}(\bbr^*)$, 
give rise in this way to random particle systems on $\bbr^*$. 

In \cite{BO}, the authors considered a particular class of $U(\infty)$-invariant measures on $H$, 
the Hua-Pickrell measures, and showed that the push-forwards of these measures to   
$\mathrm{Conf}(\bbr^*)$ give rise to random particle systems on $\bbr^*$ which are determinantal
with correlation kernels given by $K^{(\al,\bt)}(1/u,1/v)/(uv)$ (see (\ref{kernel}))
with parameters $\al,i\beta\in\bbr$.

The kernel (\ref{kernel}) is also (see \cite{BD}) a particular scaling limit 
of a kernel $K^{\al,\bt,\ga}(u,v)$
which has a similar structure but with the confluent hypergeometric functions replaced by the
hypergeometric functions $_2F_1(\al,\bt,\ga;z)$. The kernel  $K^{\al,\bt,\ga}(u,v)$
is the correlation kernel for a particle system that arises in the theory of representations of 
$U(\infty)$: role of the ergodic measures is now played by the indecomposable characters of $U(\infty)$
which are again parametrized by certain sequences together with some extra parameters
(see (1.4) in \cite{BD}).

The kernel (\ref{kernel}) also arises as the correlation kernel for a particle system 
in a similar way to $K^{\al,\bt,\ga}(u,v)$ above, but in place of irreducible representations
of $U(\infty$), we now consider irreducible (spherical) representations of
$U(\infty)\ltimes H(\infty)=\lim_{\to}G(N)$, the inductive limit of the semidirect product
$G(N)=U(N)\ltimes H(N)$, where $U(N)$ is the group of $N\times N$ unitary matrices and $H(N)$ 
denotes $N\times N$ Hermitian matrices: see \cite{BO} and references therein. 

As we will see in Section 2, the kernel (\ref{kernel}) can be obtained as a scaling limit
in unitary random matrix ensembles generated by the weight function $f(z,0)$ on the unit circle 
(given by (\ref{f0}) below) at a point of so-called Fisher-Hartwig singularity.
This singularity combines a root-type and a jump-type singularity characterized by the parameters
$\alpha$ and $\beta$, respectively.

In particular cases, the kernel (\ref{kernel}) reduces to a Bessel- and the sine-kernel which 
attracted much attention mostly because of their interest for random matrices.
If $\beta=0$ the confluent hypergeometric function reduces to Bessel
functions (see, e.g., \cite{Abr}):
\be\label{confbessel}
\phi(\mu,2\mu,2ix)=
\Gamma\left(\mu+{1\over 2}\right)e^{ix}\left(\frac x2\right)^{-\mu+{1\over 2}}J_{\mu-{1\over 2}}(x).
\ee
Therefore, we obtain from (\ref{kernel2})
\be
\label{Bessel1}
K^{(\alpha,0)}(u,v)\equiv K^{(\alpha)}_{Bessel1}(u,v)=
\frac{|u|^\alpha|v|^\alpha}{u^\alpha v^\alpha}\frac {\sqrt{uv}}{2}
\frac{J_{\alpha+\frac 12}(u)J_{\alpha-\frac 12}(v)-J_{\alpha+\frac 12}(v)J_{\alpha-\frac 12}(u)}{u-v}.
\ee
This kernel appeared in \cite{Akemann},\cite{KV},\cite{NS},\cite{WF}.

If $\alpha=0$, $\beta=0$, then (\ref{Bessel1}) reduces to the sine kernel
\be\la{sine}
K^{(0,0)}(x,y)\equiv K_{\sin}(x,y)=\frac{\sin(x-y)}{\pi(x-y)},
\ee
the most ubiquitous object of random matrix theory.

Note that the operator $K^{(\al,\bt)}$ is trace class (see Appendix), and consider the 
Fredholm determinant
\be\label{Fd}
\det(I-K^{(\al,\bt)})_{L^2(-s,s)}.
\ee

Because of the mentioned interpretation of $K^{(\al,\bt)}$  with $\al, i\bt \in \bbr$
as the correlation kernel for a particle system produced by a Hua-Pickrell measure, 
it is easy to see that the Fredholm determinant (\ref{Fd}) is the probability 
that all the $\al_j^{\pm}$ are less than $1/s$.

By a random matrix interpretation of the kernel (\ref{kernel}),
the determinant (\ref{Fd}) with $\al, i\bt \in \bbr$ gives the probability, 
in the bulk scaling limit, that the interval $(-s,s)$ with a Fisher-Hartwig
singularity at the center contains no eigenvalues of corresponding unitary random matrix ensembles. 

As noticed in \cite{WF,BD}, the determinant $\det(I-K^{(\al,\bt)})_{L^2(0,s)}$ is
related to a solution to the Painlev\'e V 
equation.\footnote{More generally, an $_2F_1$-kernel determinant is expressed \cite{BD} 
in terms of a solution to the Painlev\'e VI equation. For some asymptotic results 
which use this connection and conjectures see \cite{L}.} 

In this paper, we obtain the asymptotics of the Fredholm determinant
(\ref{Fd}) for large $s$, i.e., the large gap asymptotics.
Our main result is the following.

\noindent{\bf Theorem 1}
{\it
Let $K^{(\al,\bt)}$ be the operator with kernel (\ref{kernel}) acting on $L^2(-s,s)$, 
Then, as $s\to+\infty$,
\be
\label{Ps}
\det(I-K^{(\al,\bt)})_{L^2(-s,s)}=
\frac{\sqrt{\pi}G^2(1/2)G(1+2\alpha)}{2^{2\alpha^2}G(1+\alpha+\beta)G(1+\alpha-\beta)}
s^{-\frac 14-\alpha^2+\beta^2}
e^{-\frac{s^2}2+2\alpha s}\left[1+O\left(\frac 1s\right)\right],
\ee
where $G(x)$ is Barnes' G-function.
This expansion is uniform in compact subsets of the $\al$-half-plane $\Re\al>-1/2$ and 
of the $\bt$-plane outside neighborhoods of the points $\al\pm\bt=-1,-2,\dots$.
}

\noindent{\bf Remark 2}
Setting $\beta=0$ in (\ref{Ps}) and using a doubling formula for the G-function, we obtain 
the large $s$-asymptotics for the determinant with kernel (\ref{Bessel1}) where $\Re\al>-1/2$:
\be
\det(I-K^{(\alpha)}_{Bessel1})=\frac1{(2\pi)^{\alpha}}G(\alpha+1/2)G(\alpha+3/2)s^{-\frac 14-\alpha^2}
e^{-\frac{s^2}2+2\alpha s}\left[1+O\left(\frac 1s\right)\right].
 \ee

Setting $\alpha=\beta=0$ in (\ref{Ps}) and using the property
$2\ln G(1/2)=(1/12)\ln2-\ln\sqrt{\pi}+3\zeta'(-1)$, where $\zeta(x)$ is Riemann's zeta-function,
we reproduce the result for the sine-kernel determinant:
\be
\label{sin}
\ln \det(I-K_{\sin})=-\frac{s^2}2-\frac 14\ln s+\frac 1{12}\ln2+3\zeta'(-1)+O\left(\frac 1s\right),
\quad s\to\infty.
\ee
The first two terms in (\ref{sin}) were first found by  des Cloizeaux
and Mehta \cite{mehta}, and the full expansion
by Dyson \cite{Dyson}. The calculations in \cite{mehta, Dyson}  were not fully rigorous.
A proof for the first leading term was carried out by Widom \cite{widom1}.
The full asymptotics of the logarithmic 
derivative $(d/ds)\ln \det(I-K_{\sin})$ were proved by Deift, Its and Zhou
in \cite{DIZ}. Finally, the constant term in (\ref{sin}) was proved in \cite{Ehrhardt},
\cite{K}, \cite{DIKZ}.
\bigskip

\noindent{\bf Remark 3}
In the present paper, we address only the symmetric case of $L^2(a,b)$ such that $b=-a=s>0$.
However, one can apply our methods to consider non-symmetric cases as well.
\bigskip

In unitary random matrix ensembles at a hard edge of the spectrum (e.g., Jacobi at the edges or Laguerre
at zero) local correlations between eigenvalues are expressed in terms of the following Bessel
kernel first considered by Forrester \cite{Forr} (in an equivalent form):
\be
\label{Bessel2}
K_{Bessel2}^{(a)}(x,y)=
\frac{\sqrt{y}J'_{a}(\sqrt{y})J_a(\sqrt{x})-\sqrt{x}J'_{a}(\sqrt{x})J_{a}(\sqrt{y})}{2(x-y)}.
\ee

In particular, the distribution of the extreme eigenvalue is given 
in the scaling limit by the Fredholm determinant
$\det(I- K^{(a)}_{Bessel2})$, where $K^{(a)}_{Bessel2}$ is the trace-class operator 
on $L^2(0,s)$, $s>0$, with kernel (\ref{Bessel2}).
In Section 7, we prove the following asymptotic behavior of this determinant.

\noindent
{\bf Theorem 4.} 
{\it 
As $s\to +\infty$, we have uniformly in compact subsets of the half-plane $\Re a>-1$:
\be\la{Bessel2as}
\det(I- K^{(a)}_{Bessel2})_{L^2(0,s)}=
\tau_a s^{-a^2/4}e^{-s/4+a\sqrt{s}}\left(1+O(s^{-1/2})\right),\qquad \Re a>-1,
\ee
where
\be\la{const}
\tau_a=\frac{G(1+a)}{(2\pi)^{a/2}}.
\ee
}

In \cite{TWbessel},  Tracy and  Widom showed that the logarithmic
derivative  $(d/ds)\ln\det(I- K^{(a)}_{Bessel2})$ is expressed in terms of a solution 
to Painlev\'e V equation and used this fact to give a heuristic derivation of 
(\ref{Bessel2as}) with some constant $\tau_a$. Tracy and  Widom also conjectured
the value of $\tau_a$ given in (\ref{const}) using numerical calculations and 
comparison with the Dyson asymptotics for the sine-kernel determinants. 
(In fact, for $a=\mp 1/2$, the Bessel kernel (\ref{Bessel2}) reduces to sine-kernels 
appearing in orthogonal and symplectic ensembles of random matrices. 
The sine-kernel (\ref{sine}) appears in unitary ensembles.)
Very recently, a proof of the asymptotics (\ref{Bessel2as},\ref{const}) for the range 
of the parameter $|\Re a|<1$ was given by Ehrhardt \cite{Ebessel} using 
operator theory methods.

To prove Theorem 1, we use the approach of \cite{K}, \cite{DIKZ}, \cite{DIKairy},
where the asymptotics were computed, including the constant terms, of the sine-kernel 
and the Airy-kernel determinants. 

First, in Section 2, using results from \cite{DIKT}
we express (see Lemma 6) the Fredholm determinant (\ref{Fd}) as a scaling limit of 
Toeplitz determinants $D_n(\varphi)$ with certain 
symbols $f(e^{i\th})$ supported on an arc of the unit circle 
$\varphi\le\th<2\pi-\varphi$ with $\varphi=2s/n$, $n>s$. The continuation of these symbols into the 
complex plane has a Fisher-Hartwig singularity at $z=1$.
Theorem 1 then reduces to an asymptotic evaluation of such Toeplitz determinants for large $s$.

In Section 3 we derive a differential identity (\ref{di3}) for the logarithmic derivative 
$(d^2/d\varphi^2)\ln D_n(\varphi)$
at $0\le\varphi<\pi$ in terms of the solution to an associated
Riemann-Hilbert problem (in fact, in terms of the associated orthogonal polynomials which are given 
by this solution). In section 4, we obtain the series expansion of $D_n(\varphi)$
for $\varphi$ close to $\pi$. In Section 5, we solve the Riemann-Hilbert problem asymptotically and thus
obtain the asymptotic expression for the r.h.s. of the differential identity
(\ref{di3}), namely, we obtain the identity (\ref{di3as}) uniformly for
$2s/n<\varphi<\pi$, $n>s$, $s>s_0$, with some (large) $s_0>0$. 
Integration of the latter identity w.r.t. $\varphi$,
using the boundary condition of Section 4, gives the 
asymptotics of the determinants $D_n(\varphi)$ for any arc
with $2s/n<\varphi<\pi$, $n>s$, $s>s_0$, with some $s_0>0$, which is sufficient to prove Theorem 1.

In Section 7, we represent the Fredholm determinant with the Bessel kernel (\ref{Bessel2})
as a scaling limit of Hankel determinants related, via a general connection formula
of Theorem 2.6. of \cite{DIKT}, to the particular case of $D_n(\varphi)$ with $\bt=0$. 
The connection formula also involves the polynomials orthogonal w.r.t. $f(z)$ on the circular arc
which are represented by matrix elements of the solution to the Riemann-Hilbert problem
mentioned above.
We prove Theorem 4 by using asymptotic results on these polynomials and on $D_n(\varphi)$ 
from the previous section as well as an expansion for singular Hankel determinants from \cite{DIKT}.

\section{Connection with Toeplitz determinants}
The aim of this section is to derive an expression for (\ref{Fd}) in terms of 
Toeplitz determinants
(Lemma 6 below) and to fix notation for the rest of the paper.

Let $E_{\varphi}$ be an arc of the unit circle $C$ oriented counterclockwise:
\be\la{defE}
E_\varphi=\{e^{i\theta},\varphi\le\theta\le 2\pi-\varphi\},\qquad 0\le\varphi<\pi.
\ee
Consider the following function $f(z,\varphi)$ on $E_\varphi$:
\be
\label{weight}
f(z,\varphi)=|z-1|^{2\alpha}z^{\beta}e^{-i\pi\beta},\qquad
z=e^{i\theta} \in E_\varphi\quad \al,\bt\in\bbc,\quad \Re\al>-{1\over 2}.
\ee
Note that for $z\in C$
\be\la{zabs}
|z-1|^{2\alpha}=\frac{(z-1)^{2\alpha}}{z^{\alpha}e^{i\pi\alpha}},
\ee
where the cut of $(z-1)^{2\alpha}$ is along $[1,\infty)$, and $0<\arg(z-1)<2\pi$.
The branches of $z^\al$, $z^\bt$ are chosen so that $0<\arg z<2\pi$. Therefore, we can extend the
function $f(z)$ to the complex plane with the cut $[0,\infty)$ by the expression:
\be\la{fext}
f(z)=z^{-\al+\bt}(z-1)^{2\al}e^{-i\pi(\al+\bt)}\qquad z\in\mathbb C \setminus [0,\infty).
\ee

Related to the function (\ref{weight}) is a system of orthogonal
polynomials $p_k(z;\varphi)=\chi_k(\varphi)z^k+\dots$,
$\widehat p_k(z;\varphi)=\chi_k(\varphi)z^k+\dots$ of degree $k=0,1,\dots$, satisfying
\be
\label{p}
\frac 1{2\pi}\int\limits_\varphi^{2\pi-\varphi}p_k(z)z^{-m}f(z)d\theta=\chi_m^{-1}\delta_{km},\qquad
\frac 1{2\pi}\int\limits_\varphi^{2\pi-\varphi}\widehat p_k(z^{-1})z^{m}f(z)d\theta=\chi_m^{-1}\delta_{km},
\ee
$$ \quad z=e^{i\theta},\quad m=0,1,\dots,k. $$
Note that if the weight function $f(z)$ is not positive on $E_\varphi$, the existence
of such a system of polynomials is not a priori clear and will be addressed in the 
situations needed below.

In order to obtain the kernel (\ref{kernel}) in a scaling limit,
we will need to know the asymptotics of the polynomials
\be\label{q0}
q_n(z)\equiv p_n(z;0),\qquad \widehat q_n(z)\equiv \widehat p_n(z;0),
\ee
corresponding to the weight 
\be\la{f0}
f(z,0)=|z-1|^{2\alpha}z^{\beta}e^{-i\pi\beta},\qquad  z=e^{i\th},\qquad 0\le\th<2\pi.
\ee
In this case the function $f(z)$ possesses a Fisher-Hartwig singularity
at the point $z=1$. The asymptotics of these and more general polynomials were recently analyzed
in \cite{DIKT} (in particular, the polynomials exist for sufficiently large degrees) 
and from those results we obtain the behavior of the polynomials in a neighborhood of the
singular point.

\noindent
{\bf Lemma 5 }{\it
Let $0<\ep<1$, $U_0=\{z, |z-1|<\ep\}$. Fix the branch of $\ln z=\ln|z|+i\arg z$ by 
the condition $-\pi<\arg z<\pi$, and
the branches of the power functions $w^a=|w|^a\exp\{ia\arg w\}$  by the condition $0<\arg w<2\pi$.
Then, as $n\to\infty$, $z\in U_0$,
\begin{eqnarray}
\eqalign{
q_n(z)=
\left \{\begin {array}{c}
                  1,\quad  z\in\bbc_+\cap U_0 \\
                  e^{-2\pi i(\al-\bt)},\quad  z\in \bbc_- \cap U_0
                  \end {array}\right\}
(n\ln z)^{\al-\bt} (z-1)^{-\al+\bt}z^{\alpha-\beta}\\
\times\frac{\Gamma(1+\alpha+\beta)}{\Gamma(1+2\alpha)}
\phi(1+\alpha+\beta,1+2\alpha,n\ln z)\left[1+O\left(\frac 1n\right)\right],}\label{qn}\\
\widehat q_n(z^{-1})=(n\ln z)^{\al+\bt}(z-1)^{-(\al+\bt)}
\frac{\Gamma(1+\alpha-\beta)}{\Gamma(1+2\alpha)}
\phi(1+\alpha-\beta,1+2\alpha,-n\ln z)\left[1+O\left(\frac 1n\right)\right].\label{hqn}
\end{eqnarray}
These asymptotics are uniform and differentiable for $z\in U_0$.
They are also uniform in compact subsets of the $\al$-half-plane $\Re\al>-1/2$ and 
of the $\bt$-plane outside neighborhoods of the points $\al\pm\bt=-1,-2,\dots$.
}

\noindent
{\bf Remark}
It is easy to check that the singularities in (\ref{qn}), (\ref{hqn}) cancel.

\noindent{\it Proof.}
Our polynomials correspond to the special case of \cite{DIKT} with only one singularity located at $z=1$.
Let $U_0$ be the neighborhood of $1$ where the parametrix was constructed in  \cite{DIKT}
in terms of the confluent hypergeometric function. Let
\[
\ze= n\ln z.
\]
Take $\ze\in I$, where $I$ is the first sector of the
image of the neighborhood under the conformal transformation $\ze=n\ln z$ (see Figure 2 of \cite{DIKT}).
Tracing back the Riemann-Hilbert transformations of \cite{DIKT}, it is straightforward to obtain
\be
\pmatrix{\chi_n^{-1}q_n(z)\cr
-\chi_{n-1}z^{n-1}\widehat q_{n-1}(z^{-1})}=
\left(I+n^{-\Re\bt\si_3}O\left({1\over n}\right)n^{\Re\bt\si_3}\right)
\pmatrix{Q_1\cr Q_2},\qquad n\to\infty
\ee
where
\be
\eqalign{
Q_1(z)=\ze^{-\al-\bt} (z-1)^{-\al+\bt}z^{\alpha-\beta}e^{i\pi(\al+\bt)}\Psi_1,\\
\Psi_1=-\psi(1-\alpha+\beta,1-2\alpha,\zeta)
\frac{\Gamma(1+\alpha+\beta)}{\Gamma(\alpha-\beta)}+
\psi(-\alpha-\beta ,1-2\alpha,e^{-i\pi}\zeta)z^n,}
\ee
\be
\eqalign{
Q_2=-\ze^{\al+\bt}(z-1)^{-\alpha-\beta}e^{i\pi(\al+\bt)}\Psi_2,\\
\Psi_2=
\psi(\alpha+\beta,1+2\alpha,\zeta)-
\psi(1+\alpha-\beta ,1+2\alpha,e^{-i\pi}\zeta)
\frac{\Gamma(1+\alpha-\beta)}{\Gamma(\alpha+\beta)}e^{-2i\pi\al}z^n.\label{Psi2}}
\ee
Here $\psi(a,c,z)$ is the confluent hypergeometric function of the second kind (see, e.g., \cite{Abr}),
and $O(1/n)$ stands for a $2\times 2$ matrix with the matrix elements of that order.

Applying the following property of the confluent hypergeometric functions:
\be\label{hgprop}
\psi(a,c,z)=
\frac{\Gamma(1-c)}{\Gamma(a-c+1)}\phi(a,c,z)+\frac{\Gamma(c-1)}{\Gamma(a)}z^{1-c}\phi(a-c+1,2-c,z).
\ee
and Kummer's transformation
\be\label{Kummer}
\phi(a,c,z)=e^z\phi(c-a,c,-z)
\ee
to $\Psi_2$ in (\ref{Psi2}) gives
\be
\Psi_2
=\frac{\Gamma(1+\alpha-\beta)}{\Gamma(1+2\alpha)}e^{-i\pi(\al+\bt)}
z^n\phi(1+\al-\bt,1+2\al,-\ze),
\ee
which simplifies the expression for $Q_2$.
To simplify the formula for $Q_1$, we use (\ref{hgprop}) and (\ref{Kummer}) again.
We obtain for the combination $\Psi_1$ in $Q_1$:
\[
\Psi_1=\zeta^{2\al}\frac{\Gamma(1+\alpha+\beta)}{\Gamma(1+2\alpha)}e^{-i\pi(\al+\bt)}
\phi(1+\al+\bt,1+2\al,\ze).
\]
Now the expressions (\ref{qn}), (\ref{hqn}) of the lemma for $\ze\in I$ follow easily
(noting also that $\chi_n^2=1+O(1/n)$ by Theorem 1.8 in \cite{DIKT}).
As $q_n(z)$, $\widehat q_n(z)$ are polynomials
and the asymptotics of \cite{DIKT} hold uniformly for $z\in U_0$, the expressions for
$q_n(z)$, $\widehat q_n(z)$ extend by continuity to the whole neighborhood $U_0$ and hold there uniformly.
The uniformity properties in $\al$ and $\bt$ follow from the uniformity of the asymptotics in \cite{DIKT}.
The multiplier $e^{-2\pi i(\al-\bt)}$ for $q_n(z)$ in $\bbc_-\cap U_0$ appears because of the cut
of $z^a$ going through the neighborhood.
$\Box$

\bigskip

Let $D_n(\varphi)$ denote the Toeplitz determinant with symbol $f(z,\varphi)$:
\be\label{Dnphi}
D_n(\varphi)=\det(f_{j-k})_{j,k=0}^{n-1}=\frac1{(2\pi)^{n}n!}\int\limits_{E_\varphi}\dots\int\limits_{E_\varphi}\prod_{\scriptstyle1\le j<k\le
 n}|z_j-z_k|^2\prod_{\scriptstyle j=1}^{n} f(z_j,\varphi) {d z_j\over i z_j},
\ee
where $f_k$ are the Fourier coefficients of $f(z,\varphi)$:
$$
f_k=\frac 1{2\pi}
\int_\varphi^{2\pi-\varphi}f(e^{i\theta},\varphi)e^{-ik\theta}d\theta,\quad k=0,\pm1,\pm2,\dots
$$

Then the following lemma holds. (For $\al=\bt=0$ it reduces to the scaling limit used
by Dyson in his analysis of the sine-kernel determinant \cite{Dyson}.)

\noindent
{\bf Lemma 6.} {\it  Let $s>0$. Then
\be
\label{scalinglimit}
\det(I-K^{(\al,\bt)})=
\lim_{n\to \infty}\frac{D_n\left(\frac{2s}{n}\right)}{D_n(0)},
\ee
where $K^{(\al,\bt)}$ is the operator on $L^2(-s,s)$ with kernel (\ref{kernel}).}

\noindent {\it Proof}.
Assume first that $\al,i\bt\in\bbr$. Then, as follows, e.g., from (\ref{Dnphi}),
$D_n(\varphi)>0$ for all $n$, and therefore the polynomials $q_k(z)$, $\widehat q_k(z)$
exist for all $k$ (as follows from their determinantal representation: see, e.g., \cite{Deift}). 
By a standard argument \cite{M,Deift}, we first write the
term $\prod\limits_{j<k}|z_j-z_k|^2$ in (\ref{Dnphi}) as a
product of two Vandermonde determinants whose elements, 
by a suitable combination of the rows,
become the polynomials (\ref{q0})
$q_{k-1}(z_j)/\chi_{k-1}(0)$ and  $\widehat q_{k-1}(z^{-1}_j)/\chi_{k-1}(0)$,
$j,k=1,\dots,n$, respectively.
We obtain
\[
\prod\limits_{j<k}|z_j-z_k|^2=\prod_{j=0}^{n-1}\chi_j(0)^{-2}\times
\det\left(\sum_{\ell=0}^{n-1}\widehat q_\ell(z^{-1}_j)q_\ell(z_k)\right)_{1\leq j,k\leq n}.
\]
Using the well-known expression
$$
D_n(0)=\prod\limits_{j=0}^{n-1}\chi_j(0)^{-2},
$$
we obtain
\be
\label{DD}
\frac{D_n(\varphi)}{D_n(0)}=\frac1{(2\pi)^n n!}\int \limits_{E_\varphi}\dots \int\limits_{E_\varphi}\det\left(K_n(z_i,z_j)\right)_{1\leq i,j\leq n}{dz_1\over i z_1}\dots {dz_n\over iz_n},
\ee
where the kernel $K_n$ is given by the expression:
\be\label{CD}
\eqalign{
K_n(z_1,z_2)=\sqrt{f(z_1,0)f(z_2,0)}\sum\limits_{k=0}^{n-1}\widehat q_k(z_1^{-1})q_k(z_2)\\
=\sqrt{f(z_1,0)f(z_2,0)}\frac{(z_2/z_1)^n q_n(z_1)\widehat q_n(z_2^{-1})-\widehat q_n(z_1^{-1})q_n(z_2)}{1-z_2/z_1}.}
\ee
To obtain the second equality here, we used the Christoffel-Darboux formula:
see, e.g., Lemma 2.3. in \cite{DIKT}. 
Since for sufficiently large $n$, both the polynomials $q_n(z)$, $\widehat q_n(z)$ 
exist (see Lemma 5) and $D_n(0)\neq 0$ (see (\ref{Cn}) below) for complex $\al$ and $\bt$,
equation (\ref{DD}) is extended to the general case of $\al$, $\bt$ from $\al,i\bt\in\bbr$
by continuity and holds for all sufficiently large $n$ for $\al$ and $\bt$ in a compact set.

As, e.g., in \cite{Deift} Section 5.4.,  one shows that
the r.h.s. of (\ref{DD}) can be written as the  Fredholm determinant $\det(I-K_n)$,
where $K_n$ is the operator with  kernel (\ref{CD}) acting on
$L^2(C\setminus E_\varphi,\frac{dz}{2\pi iz})$; that is
\be\label{Dform}
D_n(\varphi)=D_n(0)\det(I-K_{n})_{L^2(C\setminus E_\varphi)},
\ee
where the arc $C\setminus E_\varphi$ is
oriented counterclockwise.

We now show that the kernel (\ref{kernel}) can be obtained as a scaling limit of (\ref{CD}).
Setting
$z=e^{\frac{2iu}{n}}$, $u>0$, in (\ref{qn}), (\ref{hqn}), we obtain
\begin{eqnarray}
q_n(e^{\frac{2iu}{n}})=
n^{\al-\bt}\frac{\Gamma(1+\alpha+\beta)}{\Gamma(1+2\alpha)}
\phi(1+\alpha+\beta,1+2\alpha,2iu)\left[1+O\left(\frac 1n\right)\right],\label{qne}\\
\widehat q_n(e^{-\frac{2iu}{n}})=
n^{\al+\bt}\frac{\Gamma(1+\alpha-\beta)}{\Gamma(1+2\alpha)}
\phi(1+\alpha-\beta,1+2\alpha,-2iu)\left[1+O\left(\frac 1n\right)\right].\label{hqne}
\end{eqnarray}
Setting now $z=e^{2\pi i+\frac{2iu}{n}}$, $u<0$, in (\ref{qn}), (\ref{hqn}),
we obtain the same expressions as the r.h.s. of (\ref{qne}), (\ref{hqne}) for the values of the polynomials
$q_n$, $\widehat q_n$ at these points.

Let $z(w)=e^{\frac{2iw}{n}}$ if $w>0$, and $z(w)=e^{2\pi i+\frac{2iw}{n}}$ if $w<0$.
Substituting the just found values of the polynomials into (\ref{CD}), we obtain
\[
\lim\limits_{n\to \infty} K_n(z(u),z(v))\frac {d z(v)}{2\pi i z(v)}
=e^{i(v-u)} K^{(\al,\bt)}(u,v)dv,\qquad u,v\in\bbr,
\qquad {dz(v)\over 2\pi i z(v)}={dv\over\pi n},
\]
where $K^{(\al,\bt)}(u,v)$ is given by (\ref{kernel}).

Using the following standard recurrence relations for the confluent hypergeometric function
\be\la{recurrence}
\eqalign{
a\phi(a+1,c,x)-(a-c+1)\phi(a,c,x)-(c-1)\phi(a,c-1,x)=0,\\
c\phi(a+1,c,x)-c\phi(a,c,x)-x\phi(a+1,c+1,x)=0,}
\ee
we can rewrite (\ref{kernel}) in the form (\ref{kernel2}).

Now using the estimates (\ref{qne})--(\ref{hqne}) for the polynomials $q_n(z)$, we see that
for any $s>0$ there exists $c(s)>0$ such that
\begin{equation}\label{estimate}
\left|\partial^{j}_{u}\partial^{k}_{v}\left(
\frac{1}{\pi n}K_{n}\left(z_1, z_2\right)  - 
e^{i(v-u)}K^{(\al,\bt)}(u,v)\right)\right|\leq \frac{c}{n},\end{equation}
where $u, v \in (-s,s)$,  $j,k = 0, 1$.
By similar arguments to those of the proof of Corollary 1.3 in \cite{DG}, the estimate
(\ref{estimate}) leads to (\ref{scalinglimit}).
$\Box$

A well-known result on asymptotics of Toeplitz determinants with a single Fisher-Hartwig singularity
such that $\al\pm\bt\neq -1,-2,\dots$
(see, e.g., \cite{Ehr, DIKT}) reads in our case:
\be
\label{Cn}
D_n(0)=n^{\alpha^2-\beta^2}\frac{G(1+\alpha+\beta)G(1+\alpha-\beta)}{G(1+2\alpha)}(1+o(1)).
\ee
This expansion is uniform \cite{DIKT} in compact subsets of the $\al$-half-plane $\Re\al>-1/2$ and 
of the $\bt$-plane outside neighborhoods of the points $\al\pm\bt=-1,-2,\dots$.
In the next sections we will obtain an expression
for $D_n(2s/n)$ for large $s$, $n$, $s<n$.

\section{Riemann-Hilbert problem and a differential identity}\la{sectionRHP-di}
Suppose that the system of orthonormal polynomials satisfying (\ref{p}) exists and
consider the following matrix-valued function:
\begin{equation} \label{Y}
    Y^{(n)}(z) =
\pmatrix{
\chi_n^{-1} p_n(z) &
\chi_n^{-1}\int_{C_\varphi}{p_n(\xi)\over \xi-z}
{f(\xi)d\xi \over 2\pi i \xi^n} \cr
\chi_{n-1}z^{n-1}\widehat p_{n-1}(z^{-1}) &
\chi_{n-1}\int_{C_\varphi}{\widehat p_{n-1}(\xi^{-1})\over \xi-z}
{f(\xi)d\xi \over 2\pi i \xi}},\qquad z\notin C_{\varphi},
\end{equation}
where $C_{\varphi}$
is the arc $E_\varphi$ (\ref{defE}) but oriented clockwise.
Denote $z_+=e^{i\varphi}$, $z_-=e^{i(2\pi-\varphi)}$, the endpoints of the arc.

It is easy to verify directly that
$Y(z)=Y^{(n)}(z)$ solves the following Riemann-Hilbert problem:
\begin{enumerate}
    \item[(a)]
        $Y(z)$ is  analytic for $z\in\bbc \setminus C_{\varphi}$.
    \item[(b)]
$Y(z)$ has continuous boundary values
$Y_{+}(z)$ as $z$ approaches the inner points of the arc $C_{\varphi}$
from the outside of the unit circle, and $Y_{-}(z)$, from the inside.
They are related by the jump condition
\begin{equation}\label{Yjump}
            Y_+(z) = Y_-(z)
            \pmatrix{
                1 & z^{-n}f(z)\cr
                0 & 1},
            \qquad\mbox{$z\in C_{\varphi}\setminus\{z_+,z_-\}$.}
        \end{equation}
    \item[(c)]
        $Y(z)$ has the following asymptotic behavior as $z\to\infty$:
        \begin{equation}\label{Yinf}
            Y(z) = \left(I+ O \left( \frac{1}{z} \right)\right)z^{n\sigma_3},
            \qquad \mbox{where $\sigma_3=\pmatrix{1&0\cr 0&-1}$.}
    \end{equation}
 \item[(d)]
 Near the endpoints of the arc,
 \begin{equation}\label{d1}
 Y(z)=O\pmatrix{
                1 & \ln|z-z_{\pm}|\cr
                1 & \ln|z-z_{\pm}|},
                \end{equation}
as $z\to z_{\pm}, \quad z\in\bbc\setminus C_{\varphi}$.
\end{enumerate}

The solution (\ref{Y}) to the RHP (a)--(d) is unique. Note first
that $\det Y(z)=1$. Indeed, from the conditions on $Y(z)$,
$\det Y(z)$ is analytic across $C_\varphi$, has all singularities
removable, and tends to $1$ as $z\to\infty$.
It is then identically $1$ by Liouville's theorem. Now
if there is another solution $\wt Y(z)$, we easily obtain by
Liouville's theorem that $Y(z) \wt Y(z)^{-1}\equiv 1$.

A general fact that orthogonal polynomials can be so represented as a solution
of a Riemann-Hilbert problem was noticed for polynomials on the real line
by Fokas, Its, Kitaev in \cite{FIK}, and extended to polynomials on the circle in \cite{BDJ}.
The point of this representation is that the Riemann-Hilbert problem can be  efficiently analyzed
for large $n$ by a steepest-descent method discovered by
Deift and Zhou \cite{DZ} (and developed further in many subsequent works).
This gives the large-$n$ asymptotics of $Y(z)$, and therefore, by (\ref{Y}),
the asymptotics of the orthogonal polynomials. We defer the asymptotic analysis in the present case
to Section \ref{sectionRHasymptotics}. The rest of this section will be devoted to
a derivation of a differential identity for $D_n(\varphi)$ in terms of the matrix elements
of $Y(z)$.

We start with the following auxiliary lemma (which, in fact, is true for any weight $f(z)$
and a jump contour $C_\varphi$).

\noindent
{\bf Lemma 7.} {\it Let the system of polynomials $p_k(z)$, $\widehat p_k(z)$, $k=0,\dots$ satisfying
(\ref{p}) exist. Fix $n\ge 1$.
Then we have the following Christoffel-Darboux identity in terms of
the function (\ref{Y}):
\be\la{CD2}
\sum_{k=0}^{n-1}\widehat p_k(z^{-1})p_k(z)=
-z^{-n+1}\lim_{\ze\to z}
\tr\left({dY(\ze)\over d\ze}\pmatrix{0 & 1\cr 0 & 0}Y(\ze)^{-1}\right),\qquad \ze\notin C_\varphi.
\ee
}

\noindent {\bf Remark.} The right hand side of (\ref{CD2}) contains only the elements of the first
column of $Y$ which are analytic.

\noindent{\it Proof.}
Multiplying the recurrence relation (2.4) of Lemma 2.1. in \cite{DIKT} by $z$, and replacing
$n$ with $n-1$ gives (in the present notation)
\[
z\widehat p_n(z^{-1})=
{\chi_{n-1}\over\chi_n}\widehat p_{n-1}(z^{-1})+
{\widehat p_n(0)\over \chi_n}z^{-n+1}p_n(z).
\]
Substituting this expression into the r.h.s. of the Christoffel-Darboux identity (2.8) of \cite{DIKT}:
\[
\sum_{k=0}^{n-1}\widehat p_k(z^{-1})p_k(z)=
-n p_n(z)\widehat p_n(z^{-1}) +z\left(\widehat p_n(z^{-1}){d\over dz} p_n(z)-
p_n(z){d\over dz}\widehat p_n(z^{-1})\right),
\]
we obtain
\be\la{cdint}\eqalign{
\sum_{k=0}^{n-1}\widehat p_k(z^{-1})p_k(z)=
-(n-1){\chi_{n-1}\over\chi_n}z^{-1} p_n(z)\widehat p_{n-1}(z^{-1})\\
-{\chi_{n-1}\over\chi_n}\left[p_n(z){d\over dz}\widehat p_{n-1}(z^{-1})-
\widehat p_{n-1}(z^{-1}){d\over dz}p_n(z)\right].}
\ee
Using the expressions
$Y_{11}(z)=\chi_n^{-1}p_n(z)$, $Y_{21}=\chi_{n-1}z^{n-1}\widehat p_{n-1}(z^{-1})$, and
$\chi_{n-1}{d\over dz}\widehat p_{n-1}(z^{-1})=-(n-1)z^{-n}Y_{21}(z)+z^{-n+1}{d\over dz}Y_{21}(z)$,
we obtain from (\ref{cdint})
\be
\sum_{k=0}^{n-1}\widehat p_k(z^{-1})p_k(z)=
z^{-n+1}(Y_{21}{d\over dz}Y_{11}-Y_{11}{d\over dz}Y_{21}),
\ee
which proves the Lemma. $\Box$.

We are now ready to formulate the main result of this section.

\noindent
{\bf Lemma 8.} {\it  Let the polynomials $p_k(z)$, $\widehat p_k(z)$, $k=N_0,N_0+1,\dots$, satisfying
(\ref{p}) exist for some $N_0\ge 0$. Fix $n>N_0$. Let
$z_{+}$, $z_{-}$ be the endpoints of the arc $C_\varphi$:
$z_{+}=e^{i\varphi}$, $z_{-}=e^{i(2\pi-\varphi)}$.
Then
\be\la{di2}
\frac{1}{i}\frac{d}{d\varphi}\ln D_n(\varphi)=
\tr C_0(C_+-C_-)+\frac{z_+}{z_+-1}\tr C_1 C_+ -\frac{z_-}{z_--1}\tr C_1 C_-
+\frac{z_++z_-}{z_+-z_-}\tr C_+C_-,
\ee
where, in terms of the matrix (\ref{Y}),
\be\la{C01}
C_0=-\frac{n+\alpha-\beta}2Y(0)\sigma_3Y^{-1}(0),\qquad
C_1=\alpha Y(1)\sigma_3 Y^{-1}(1),
\ee
and the following limits taken for $z\notin C_\varphi$ exist and define $C_\pm$:
\be\la{Cpm}
C_+=\lim_{z\to z_+}\frac{z^{-n}f(z)}{2\pi i}
Y(z)\pmatrix{ 0 & 1 \cr 0 & 0} Y^{-1}(z),\qquad
C_-=-\lim_{z\to z_-}\frac{z^{-n}f(z)}{2\pi i}
Y(z)\pmatrix{ 0 & 1 \cr 0 & 0} Y^{-1}(z),
\ee
with the extension of $f$ given by (\ref{fext}). Moreover, the second derivative
\be\la{di3}
\frac{d^2}{d\varphi^2}\ln D_n(\varphi)=
\frac{z_+}{(z_+-1)^2}\tr C_1(C_+ + C_-)
+\frac{4}{(z_+-z_-)^2}\tr C_+C_-.
\ee
}

\noindent
{\bf Proof.}
Assume first that $\al,i\bt\in\bbr$. Then we can set $N_0=0$ (cf. proof of Lemma 6).
Starting with the representation of a Toeplitz determinant in terms of the
leading coefficients of the polynomials $p_k$:
\be
D_n(\varphi)=\prod\limits_{k=0}^{n-1}\chi_k^{-2}(\varphi),
\ee
we obtain, using first (\ref{p}) and then integration by parts, that
\be\eqalign{
\frac d{d\varphi}\ln D_n(\varphi)=-2\sum\limits_
{k=0}^{n-1}\frac{\chi_k'(\varphi)}{\chi_k(\varphi)}
=-\frac 1{2\pi}\int_{\varphi}^{2\pi-\varphi}\sum\limits_{k=0}^{n-1}
\frac\partial{\partial
\varphi}\left(p_k(z)\widehat{p}_k(z^{-1})\right)f(z,\varphi)d\theta=\\
-\frac 1{2\pi}\sum_{k=0}^{n-1}
\left[
p_k(z_{-})\widehat{p}_k(z_{-}^{-1})f(z_{-},\varphi)
+p_k(z_{+})\widehat{p}_k(z_{+}^{-1})f(z_{+},\varphi)\right].}
\ee
The previous lemma immediately gives
\be\la{di1}\eqalign{
2\pi\frac d{d\varphi}\ln D_n(\varphi)=
z_+^{-n+1}f(z_+,\varphi)\lim_{z\to z_+}
\tr\left({d Y\over dz}(z)\pmatrix{0 & 1\cr 0 & 0}Y(z)^{-1}\right)\\
+z_-^{-n+1}f(z_-,\varphi)\lim_{z\to z_-}
\tr\left({d Y\over dz}(z)\pmatrix{0 & 1\cr 0 & 0}Y(z)^{-1}\right).}
\ee
In the next section we will show that the solution $Y$ of the Riemann-Hilbert problem
exists for all $n>N_0$ with $N_0$ sufficiently large uniformly for $\al$ and $\bt$ in a compact set.
Therefore, the identity (\ref{di1}) extends to the general complex $\al$ and $\bt$ from
$\al,i\bt\in\bbr$ by continuity.   

One could already use the identity (\ref{di1}) for the purposes of the present paper. However,
following the philosophy of the Riemann-Hilbert-problem approach \cite{DIZ},
we can simplify it further to the form (\ref{di2}) which does not contain derivatives of $Y$.
In order to do this, consider the function
 \be
 \label{defwtY}
 \wt Y(z)=Y(z)\om(z)^{\si_3/2},\qquad \om(z)=z^{-n} f(z),
 \ee
where $f$ outside the arc is given by (\ref{fext}).
The function $\wt Y(z)$ is easily seen to be the solution of the problem:
\begin{enumerate}
    \item[(a)]
        $\wt Y(z)$ is analytic for $z\in\bbc \setminus (C_\varphi\cup \bbr_+)$;
    \item[(b)] On the contours $C_\varphi$ and $\bbr_+$:
\be
\wt Y_+(z)=\wt Y_-(z) \pmatrix{ 1 & 1 \cr 0 & 1},\quad z\in C_\varphi\setminus\{z_+,z_-\};\la{b2}
\ee
\be
\wt Y_+(z)=\wt Y_-(z) \left({\om_+\over\om_-}\right)^{\si_3/2},\quad z\in (0,+\infty);
\ee
    \item[(c)]
  $\wt Y(z)=\left(I+O(\frac {1}{z})\right)z^{\frac{n\sigma_3}2}
f(z)^{\frac{\sigma_3}2},\quad z\to\infty $.
\end{enumerate}

Since ${\om_+\over\om_-}$ is constant on $(0,+\infty)$, we see that
\[
\wt F(z)\equiv{d\wt Y\over dz}\wt Y^{-1}
\]
has no jumps. Since
\be\la{om}
{d\om\over dz}(z)=
\left(-{n\over z}-{\al-\bt\over z}+{2\al\over z-1}\right)\om(z),\qquad
{d\over dz}\left(\om(z)^{\si_3}\right)=\si_3{\om'(z)\over\om(z)}\om(z)^{\si_3},
\ee
we obtain from the condition (c) for $\wt Y$ that
$\wt F(z)=O(1/z)$ as $z\to\infty$. The Riemann-Hilbert problem shows that
this function can have isolated singularities at $z_+$, $z_-$, $0$, and $1$.
First, we obtain using (\ref{om}):
\be
\wt F(z)=\frac{C_0}{z}+T_0(z),\quad z\to 0;\qquad
\wt F(z)=\frac{C_1}{z-1}+T_1(z),\quad z\to 1,
\ee
where $C_0$, $C_1$ are given by (\ref{C01}), and $T_j(z)$ are Taylor series.

Now let $U$ be a disk of a sufficiently small radius centered at $z_+$.
If $\widehat Y$ is defined by the expression
\be\la{defhatY}
\wt Y(z)=\widehat Y(z)\pmatrix{1& {1\over 2\pi i}\ln(z-z_+)\cr 0 & 1},\qquad z\in U,
\ee
then it follows from (\ref{b2}) that $\widehat Y(z)$ in $U$ has no jump, and
from (\ref{d1}), that its singularity at $z_+$ is removable. Thus
$\widehat Y(z)$ is analytic in $U$.
Using (\ref{defhatY}) we then obtain
\be
\wt F(z)=\frac{C_+}{z-z_+}+T_3(z),\quad z\to z_+;
\ee
where $T_3(z)$ is a Taylor series and
\be
C_+=\frac{1}{2\pi i}
\widehat Y(z_+)\pmatrix{ 0 & 1 \cr 0 & 0} \widehat Y^{-1}(z_+).
\ee
Using the definitions (\ref{defhatY}) and (\ref{defwtY}), we obtain the expression
(\ref{Cpm}) for $C_+$. Note that the limit in (\ref{Cpm}) exists as the logarithmic singularity
of $Y(z)$ at $z_+$ cancels from that expression.

A similar analysis at $z_-$ gives that
\be
\wt F(z)=\frac{C_-}{z-z_-}+T_4(z),\quad z\to z_-;
\ee
where $T_4(z)$ is again a Taylor series and $C_-$ is defined in (\ref{Cpm}).

Thus we conclude that $\wt F(z)$ is a meromorphic function with first-order poles
at $0$, $1$, $z_+$, $z_-$, and since $\wt F(z)=o(1)$ at infinity, we have
identically in the complex plane
\[
\wt F(z)=\frac{C_0}{z}+\frac{C_1}{z-1}+
\frac{C_+}{z-z_+}+\frac{C_-}{z-z_-},
\]
or recalling the definitions of $\wt F(z)$ and $\wt Y(z)$,
\begin{eqnarray}\la{Yder0}
{d\wt Y\over dz}(z,\varphi)=
\left(\frac{C_0}{z}+\frac{C_1}{z-1}+
\frac{C_+}{z-z_+}+\frac{C_-}{z-z_-}\right)\wt Y(z,\varphi),\\
\la{Yder}
{d Y\over dz}(z,\varphi)=
\left(\frac{C_0}{z}+\frac{C_1}{z-1}+
\frac{C_+}{z-z_+}+\frac{C_-}{z-z_-}\right)Y(z,\varphi)-
\frac{\om'(z)}{2\om(z)}Y(z,\varphi)\si_3.
\end{eqnarray}
Substituting  (\ref{Yder}) into (\ref{di1}) and noticing that $C_\pm^2=0$
gives the identity (\ref{di2}).

To obtain the identity for the second derivative, note first that
as follows from the general theory the function $\wt Y(z)$ is differentiable
w.r.t. $\varphi$. Similarly to our derivation of (\ref{Yder0}), we obtain
\be\la{Yderp}
{d\wt Y\over d\varphi}(z,\varphi)=
\left(
\frac{-iz_+C_+}{z-z_+}+\frac{iz_-C_-}{z-z_-}\right)\wt Y(z,\varphi).
\ee
Equating the derivatives ${d\over dz}{d\over d\varphi} \wt Y(z,\varphi)=
{d\over d\varphi}{d\over dz} \wt Y(z,\varphi)$, gives by (\ref{Yder0}) and (\ref{Yderp})
a compatibility condition on $C_0$, $C_1$, $C_+$, $C_-$, and their derivatives w.r.t. $\varphi$.
Equating the coefficients at $1/z$ in this condition gives
\be
{d\over d\varphi}C_0 +i[C_0,C_+ - C_-]=0,
\ee
where $[A,B]=AB-BA$.
Similarly the coefficients at
$1/(z-1)$, $1/(z-z_+)$, $1/(z-z_-)$ yield the identities:
\begin{eqnarray}
{d\over d\varphi}C_1 +i\left[C_1,{z_+\over z_+ -1}C_+ -{z_-\over z_- -1}C_-\right]=0,\\
{d\over d\varphi}C_+ -i[C_0,C_+]-{iz_+\over z_+ -1}[C_1,C_+]+
i{z_+ + z_-\over z_+ - z_-}[C_+,C_-]=0,\\
{d\over d\varphi}C_- +i[C_0,C_-]+{iz_-\over z_- -1}[C_1,C_-]
-i{z_+ + z_-\over z_+ - z_-}[C_+,C_-]=0.
\end{eqnarray}

Differentiating (\ref{di2}) w.r.t. $\varphi$ and substituting the above identities for the derivatives of
$C_0$, $C_1$, $C_+$, $C_-$ in the resulting expression gives the formula (\ref{di3}).
In this calculation, it is convenient to use the elementary algebraic identity:
\be\la{ABC}
\tr [A,B]C= \tr A[B,C],
\ee
which equals $0$ if any two of $A$, $B$, $C$ coincide.
$\Box$

\section{Expansion of $D_n(\varphi)$ as $\varphi\to \pi$.}\la{sectionDatpi}
For a fixed $n\ge 1$ we will now obtain the expansion of $D_n(\varphi)$ as  $\varphi\to\pi$.
We use the representation (\ref{Dnphi}) of $D_n(\varphi)$ as a multiple integral:
\[
 D_n(\varphi)=\frac1{(2\pi)^{n}n!}\int_{\varphi}^{2\pi-\varphi}\cdots\int_{\varphi}^{2\pi-\varphi}
\prod_{1\le j<k\le n}|e^{i\theta_j}-e^{i\theta_k}|^2\prod_{j=1}^{n}
|e^{i\theta_j}-1|^{2\alpha}e^{i\theta_j\beta}e^{-i\pi\beta} d\theta_j.
\]

For the analysis of $D_n(\varphi)$ as $\varphi\to \pi$,
set $\varphi=\pi-\varepsilon,\quad \varepsilon>0$.
Substituting $\theta_j=\pi+\varepsilon x_j$ in the integrals, we obtain:
\be\eqalign{\la{58}
D_n(\varphi)=\frac{\varepsilon^n}{(2\pi)^{n}n!}
\int_{-1}^{1}\cdots\int_{-1}^{1}\prod_{1\le j<k\le
 n}|e^{i\varepsilon x_j}-e^{i\varepsilon x_k}|^2\prod_{j=1}^{n}
|e^{i\varepsilon x_j}+1|^{2\alpha}e^{i\varepsilon x_j\beta}dx_j\\
=\frac{\varepsilon^{n^2}2^{2\alpha n}}{(2\pi)^n}\left(A_n+O(\varepsilon^2)\right),\qquad
\mbox{as $\ep\to 0$, $n$ fixed},}
\ee
where
\be \label{An}
A_n={1\over n!}
\int_{-1}^{1}\cdots\int_{-1}^{1}
\prod_{1\leq j < k \leq n}(x_j - x_k)^{2}
\prod_{j=1}^n dx_j=2^{n^2}\prod_{k=0}^{n-1}
\frac{k!^3}{(n+k)!}
\ee
is a Selberg integral (or a product $\prod_{k=0}^{n-1}\varkappa_k^{-2}$, where
$\varkappa_k$ are the leading coefficients of the orthonormal
Legendre polynomials). The error term in (\ref{58}) is of order $\varepsilon^2$.
Indeed, it is easy to see that the expansion of the factors
with the absolute value in the integrand in (\ref{58}) gives an error of order
$\varepsilon^2$. The factors $e^{i\varepsilon x_j\bt}$ produce the following term
of order $\varepsilon$:
\[
{i\varepsilon\bt\over n!}
 \int_{-1}^{1}\cdots\int_{-1}^{1}\sum_{j=1}^n x_j
\prod_{1\leq j < k \leq n}(x_j - x_k)^{2}
\prod_{j=1}^n dx_j,
\]
which is equal to $0$, as the change of variables $x_j\rightarrow -x_j$ shows.
Therefore, the error term in (\ref{58}) is indeed $O(\varepsilon^2)$.

The asymptotics of $A_n$ as $n\to\infty$ are \cite{Widom}:
\[
\ln A_n=-n^2\ln 2 +n\ln(2\pi) -{1\over 4}\ln n + \frac{1}{12}\ln 2
+ 3\zeta'(-1)+o(1) ,\qquad n \to \infty,
\]
where $\zeta'(z)$ is the derivative of Riemann's zeta-function.
Therefore,
\be
\label{Dnpi}
\ln D_n(\varphi)=n^2\ln(\pi-\varphi)+(2\alpha n-n^2)\ln 2-\frac 14\ln n+\frac 1{12}\ln 2+3\zeta'(-1)
+\delta_n+O_n(\varepsilon^2),\quad \varepsilon=\pi-\varphi.
\ee
Here $\delta_n$ depends only on $n$ and $\delta_n\to 0$ as $n\to \infty$. The term
$O_n(\varepsilon^2)\to0$, as $\varepsilon\to 0$, $n$ fixed.

\section{Asymptotic analysis of $Y(z)$}\la{sectionRHasymptotics}
We now analyze the Riemann-Hilbert (RH) problem of Section \ref{sectionRHP-di}
for $Y(z)$ in the limit of large $n$.
The analysis is similar to that of \cite{K} and \cite{DIKZ}. Consider the function
\be
\label{psi}
\Psi(z)=\frac 1 {2\gamma}\left(z+1+\sqrt{(z-z_{+})(z-z_{-})}\right), \qquad \gamma=\cos(\varphi/2),
\ee
which conformally maps the outside of the arc $C_{\varphi}$ onto the outside of the unit circle.
Note that $\Psi_+\Psi_-=z$ for $z\in C_\varphi\setminus\{z_+,z_-\}$. Furthermore, we see as in
\cite{K} that
\be\la{ineq}
\left|{z\over\Psi(z)^2}\right|<1,\qquad \mbox{for $|z|\neq 1$}.
\ee

We apply several transformations to the Riemann-Hilbert problem. First, set
\be
\label{T}
T(z)=\gamma^{-n\sigma_3}Y(z)\Psi(z)^{-n\sigma_3}.
\ee
Then we obtain a RH problem which is normalized to $I$ at infinity:
\begin{enumerate}
    \item[(a)]
        $T(z)$ is analytic for $z\in\bbc \setminus C_{\varphi}$.
    \item[(b)] $T(z)$ has $L^2$ boundary values on $C_\varphi$ related by the condition
\begin{equation}\label{Tjump}
            T_+(z) = T_-(z)
            \pmatrix{
                z^n \Psi_+(z)^{-2n} & f(z)\cr
                0 & z^n \Psi_-(z)^{-2n}},
            \qquad\mbox{for $z\in C_{\varphi}\setminus\{z_+,z_-\}$.}
        \end{equation}
    \item[(c)]
                \begin{equation}\label{Tinf}
            T(z) = I+ O \left( \frac{1}{z} \right),\qquad
\mbox{as $z\to\infty$. }
    \end{equation}
 \end{enumerate}

As in \cite{DIKZ}, we now go over to the variable $\lb$ given by the following
linear-fractional transformation that maps the arc $C_\varphi$ onto the interval $[-1,1]$
with the point $z=z_-$ corresponding to $\lb=-1$, and $z=z_+$, to $\lb=1$:
\be
\label{map}
\lambda=\frac{z+1}{z-1}i\tan{\varphi\over2}, \qquad
z=\frac{\lb+i\tan\frac \varphi 2}{\lb-i\tan\frac \varphi 2}.
\ee
The complementary to $C_\varphi$ arc of the unit circle is mapped to $\mathbb R\setminus [-1,1]$.
The points $z=0$, $1$, $\infty$ are mapped to
$\lambda=-i\tan\frac \varphi 2$, $\infty$, $i \tan \frac \varphi 2$, respectively.
The cut of the function $f(z)$, $(0,1)\cup (1,+\infty)$ becomes
$(-i\tan\frac \varphi 2,-i\infty) \cup (i\infty,i\tan\frac \varphi 2)$ in the $\lambda$-plane
(see Figure 1).

\begin{figure}
\centerline{\psfig{file=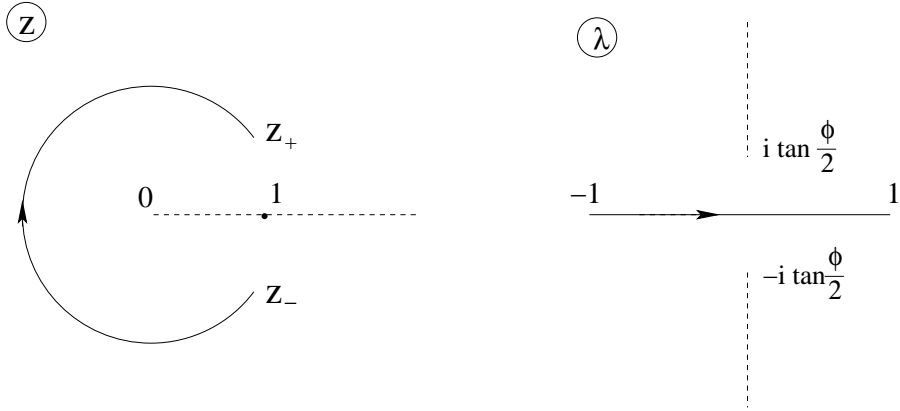,width=4.7in,angle=0}}
\vspace{0.2 cm}
\caption{
Conformal mapping.}
\label{fig1}
\end{figure}

For the case of a varying arc when $\varphi\to 0$ and $\varphi\to\pi$, in the RH analysis,
one would need to consider contracting neighborhoods of the end points $z_\pm$ in the $z$-plane
(cf. \cite{K}). This could be carried out. However,
in the $\lb$-plane, we can keep neighborhoods of the points $\lb=\pm1$ fixed, which
considerably simplifies the calculations below. Thus, going over to $\lb$ is not essential, but useful.

The problem for $T$ corresponds to the following one in the $\lb$-plane:
\begin{enumerate}
    \item[(a)]
        $\wt T(\lb)$ is  analytic for $\lb\in\bbc \setminus [-1,1]$.
    \item[(b)] The boundary values of $\wt T(\lb)$ on $(-1,1)$ are related by the condition
\begin{equation}\label{Tjump2}
            \wt T_+(\lb) = \wt T_-(\lb)
            \pmatrix{
                \Phi_+(\lb)^{-2n} & f(z(\lb))\cr
                0 & \Phi_-(\lb)^{-2n}},
            \qquad\mbox{for $\lb\in(-1,1)$,}\qquad \Phi(\lb)={\Psi(z(\lb))\over z(\lb)^{1/2}}.
        \end{equation}
    \item[(c)]
                \begin{equation}\label{Tinf2}
            \wt T(\lb) = I+ O \left( \frac{1}{\lb} \right),\qquad
\mbox{as $\lb\to\infty$. }
    \end{equation}
 \end{enumerate}

The solution $\wt T(\lb)$ is related to $T(z)$ by the expression:
\be\label{Ttilde}
T(z)=T_0 \wt T(\lb(z)),\qquad T_0=\wt T^{-1}\left(i\tan{\varphi\over 2}\right).
\ee

For the function (\ref{fext}) we have in the $\lb$-variable (we denote it $f(\lb)$ for simplicity):
\be\la{777}
f(\lb)\equiv f(z(\lb))=e^{-i\pi\bt}\left(2\tan{\varphi\over 2}\right)^{2\al}
\left(\lb-i\tan{\varphi\over 2}\right)^{-\al-\bt}
\left(\lb+i\tan{\varphi\over 2}\right)^{-\al+\bt}.
\ee

The function $\Phi(\lb)$:
\be\la{Phi}
\Phi(\lb)=\frac{\lb+i\sin{\varphi\over2}\sqrt{1-\lb^2}}{\cos{\varphi\over2}
(\lb^2+\tan^2{\varphi\over2})^{1/2}}.
\ee
Note that, because of the properties of $\Psi(z)$ discussed above, we have
\be\la{ineq2}
|\Phi(\lb)|>1,\quad \lb\notin\mathbb R;\qquad |\Phi_\pm(\lb)|=1,\quad \lb\in [-1,1].
\ee

\begin{figure}
\centerline{\psfig{file=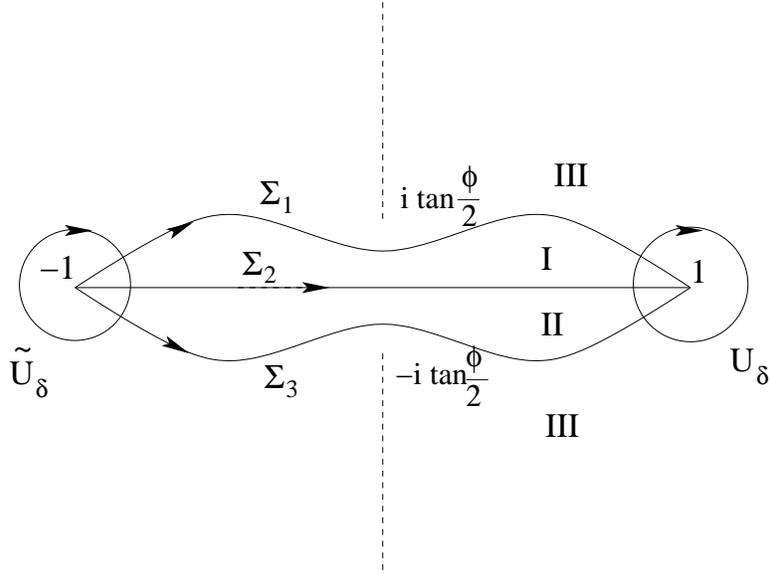,width=4.0in,angle=0}}
\vspace{0.2 cm}
\caption{
Contour for Riemann-Hilbert problems.}
\label{fig2}
\end{figure}

Following the steepest-descent method of Deift and Zhou \cite{DZ} we now change the
RH problem so that the oscillating
behavior of the matrix elements in (\ref{Tjump2})
is converted into the exponential decay as $n\to\infty$.

Namely, consider the system of contours shown in Figure 2. Let $I$ be the region bounded
by the curves $\Si_1$ and $\Si_2\equiv (-1,1)$;  region $II$ is the one bounded by the curves
$\Si_2$ and $\Si_3$; region $III$ is the rest of the complex plane.
Define the function $S(\lb)$ as follows:

\noindent
in region $I$,
\be\la{S1}
S(\lb)=\wt T(\lb)\pmatrix{1&0\cr -f(\lb)^{-1}\Phi(\lb)^{-2n}& 1},
\ee
in region $II$,
\be\la{S2}
S(\lb)=\wt T(\lb)\pmatrix{1&0\cr f(\lb)^{-1}\Phi(\lb)^{-2n}& 1},
\ee
in region $III$,
\be\la{S3}
S(\lb)=\wt T(\lb).
\ee

The Riemann-Hilbert problem for $S$ is then the following:
\begin{enumerate}
\item[(a,b)]
$S(\lb)$ is analytic in $\bbc\setminus(\Sigma_1\cup\Sigma_2\cup\Sigma_3)$ with the
following jump conditions on the contours:
\begin{equation}
            S_+(\lb) = S_-(\lb)
            \pmatrix{1&0\cr f(\lb)^{-1}\Phi(\lb)^{-2n}& 1},
\qquad\mbox{$\lb \in \Sigma_1\cup\Sigma_3$,}
\end{equation}
\begin{equation}
S_+(\lb) = S_-(\lb)
            \pmatrix{0&f(\lb)\cr
              -f(\lb)^{-1}&0},
\qquad\mbox{$\lb \in \Sigma_2\equiv (-1,1)$.}
\end{equation}
\item[(c)]
As $\lb\to\infty$,
\be
 S(\lb) = I+ O \left( \frac{1}{\lb} \right).
\ee
\end{enumerate}

For $S$ to have these properties,
the contours $\Si_{1,3}$ should not intersect the real axis and the cuts 
$(-i\infty,-i\tan\varphi/2)$, $(i\tan\varphi/2,i\infty)$ of $f(\lambda)$.

Below we will investigate the inequality in (\ref{ineq2}) in more detail and
will show that $\Si_{1,3}$ can be chosen so that
for $\varphi$ satisfying $2s/n <\varphi<\pi$, $n>s$, $s>s_0$,
the jump matrix on $\Sigma_1\cup\Sigma_3$ is uniformly
close to the identity up to an error of order $e^{-\ep s_0}$, $\ep>0$,
outside neighborhoods of the
endpoints of the arc. The error is small for $s_0$ sufficiently large.
This suggests that outside some $\de$-neighborhoods $U_\de$, $\wt U_\de$
of the endpoints, the function
$S$ can be approximated by a parametrix which has a jump only on $\Sigma_2$.
The problem for this parametrix in the outside-the-neighborhoods region is standard \cite{Deift}
and will be presented below.
It is solved explicitly. Then we will consider the neighborhoods $U_\de$, $\wt U_\de$,
and construct (following \cite{KVA}) local parametrices there in terms of Bessel functions.
We then match the outside and the local parametrices on the boundaries $\partial U_\de$,
$\partial \wt U_\de$ for large $n$, which produces the asymptotic expansion of $S$
in the inverse powers of $n\sin\frac\varphi 2$. The latter expression is large provided again
that $2s/n <\varphi<\pi$, $n>s$, $s>s_0$, and $s_0$ is sufficiently large.

\subsection{Outside parametrix}
The parametrix outside neighborhoods of $\lb=\pm 1$ is
the solution to the following RH problem:
\begin{enumerate}
    \item[(a)]
        $N(\lb)$ is  analytic for $\lb\in\bbc \setminus [-1,1]$,
    \item[(b)]
$N(\lb)$ has $L^2$ boundary values $N_+$, $N_-$ on $(-1,1)$ related as follows:
\be
N_+(\lb) = N_-(\lb)
            \pmatrix{0&f(\lb)\cr
              -f(\lb)^{-1}&0},
\qquad\mbox{$\lb \in (-1,1)$},
\ee
\item[(c)]
\be
N(\lb) = I+ O \left( \frac{1}{\lb} \right),
     \qquad \mbox{as $\lb\to\infty$.}
\ee
\end{enumerate}

As is easy to verify, this problem has the following solution:

\be
\label{N}
N(\lb)=\frac 12({\cal D}_\infty)^{\sigma_3}\pmatrix{
                a+a^{-1} & -i(a-a^{-1})\cr
                i(a-a^{-1})& a+a^{-1}}{\cal D}(\lb)^{-\sigma_3},\qquad
 a(\lb)=\left(\frac{\lb-1}{\lb+1}\right)^{1/4},
\ee
where the branch of the root is chosen so that $a(\lb)\to 1$ as $\lb\to\infty$.
The Szeg\H{o} function ${\cal D}(\lb)$ is the solution to the following RH conditions:
a) ${\cal D}(\lb)$
is analytic in $\mathbb C\setminus [-1,1]$; b) ${\cal D}_+(\lb){\cal D}_-(\lb)=f(\lb)$ for
$\lb\in (-1,1)$; c)  ${\cal D}(\lb)\to \mbox{const}$ as $\lb\to\infty$. We have
\be
\label{D}
{\cal D}(\lambda)=\exp\left(
\frac{\sqrt{1-\lambda^2}}{2\pi i}\int_{-1}^1\frac{\ln f(\eta)}{\sqrt{1-\eta^2}}\frac{d\eta}{\eta-\lambda}
\right),
\ee
with the integration over the upper (``+'') side of the interval $(-1,1)$:
we choose $\sqrt{x}>0$ for $x>0$ and $0<\arg(\lb\pm 1)<2\pi$.
Finally,
\be
\label{Dinfty}
{\cal D}_{\infty}=\lim\limits_{\lb\to \infty}{\cal D}(\lb)=
\exp\left(
\frac{1}{2\pi}\int_{-1}^1\frac{\ln f(\eta)d\eta}{\sqrt{1-\eta^2}}
\right).
\ee

In what follows, we will need an expansion of $D(\lb)$ at the endpoints $\pm 1$.
The integral in (\ref{D}) can be written as half the integral around a loop encircling $[-1,1]$.
Deforming the loop, we obtain
\be\label{ints}
\int_{-1}^1\frac{\ln f(\eta)}{\sqrt{1-\eta^2}}\frac{d\eta}{\eta-\lambda}=
\frac{\pi i}{(\lambda^2-1)^{1/2}}i\ln f(\lb) +
$$
$$
\pi i\lim_{R\to\infty}\left[(\alpha+\beta)\int_{iR}^{i\tan\frac\varphi 2} \frac{id\eta}{(\eta^2-1)^{1/2}}
\frac 1{\eta-\lambda}+(\al-\bt)\int_{-iR}^{-i\tan\frac \varphi 2}
\frac{i d\eta}{(\eta^2-1)^{1/2}}\frac 1{\eta-\lambda}\right],
\ee
where $-i(x^2-1)^{1/2}=\sqrt{1-x^2}>0$ on the upper side of $(-1,1)$.
Expanding $\frac 1{\eta-\lambda}$ near $\lambda=\pm1$, we obtain:
\begin{eqnarray}\label{ints1}
\int_{iR}^{i\tan\frac\varphi 2}\frac{id\eta}{(\eta^2-1)^{1/2}}
\frac 1{\eta-\lambda}=\pm i -e^{\pm i\varphi/2}+O(\lambda\mp 1), \quad \lambda\to \pm1;\\
\label{ints2}
\int_{-iR}^{-i\tan\frac\varphi 2}\frac{id\eta}{(\eta^2-1)^{1/2}}
\frac 1{\eta-\lambda}=\pm i +e^{\mp i\varphi/2}+O(\lambda\mp 1), \quad \lambda\to \pm1.
\end{eqnarray}

Substituting (\ref{ints1}), (\ref{ints2}) into (\ref{ints}), we obtain for (\ref{D})
\be
\label{Dexp}
\eqalign{
{\cal D}(\lb)=f(1)^{1/2}\exp\left({\cal D}_1(1)
\left(\frac{\lb-1}{2}\right)^{1/2}+
{\cal D}_2(1)\left(\frac{\lb-1}{2}\right)^{3/2}+
O\left(\lb-1\right)^{5/2}\right),
\qquad \lb\to 1,\\
{\cal D}(\lb)=f(-1)^{1/2}\exp\left({\cal D}_1(-1)
\left(\frac{\lb+1}{2e^{i\pi}}\right)^{1/2}+
{\cal D}_2(-1)\left(\frac{\lb+1}{2e^{i\pi}}\right)^{3/2}+
O\left(\lb+1\right)^{5/2}\right),\\
\lb\to -1,}
\ee
where
\be\la{D1}
{\cal D}_1(\pm 1)=2
\left(\alpha\left(1-\sin\frac \varphi 2\right)\pm i\beta \cos\frac \varphi 2  \right),
\ee
and the exact value of ${\cal D}_2(\pm 1)$ will not be used as it cancels from the final
expressions below.
It is clear from the construction that this expansion
is uniform in $\varphi$ as well as in $\al$, $\bt$ in a compact set.

\subsection{Local parametrices}\la{local}

Let $U_{\delta}$ and $\wt U_{\delta}$  denote the (nonintersecting) $\de$-neighborhoods
of the points $1$ and $-1$, respectively: see Figure 2. We choose $\de$ to be sufficiently small,
see below.

We will now write down essentially known (see \cite{DIZ,KVA,K}) parametrices $P$, and $\wt P$
in $U_{\delta}$ and $\wt U_{\delta}$, respectively. The parametrices
have the same jumps as $S$ inside these neighborhoods and match $N$ on the boundaries
$\partial U_{\delta}$ and $\partial\wt U_{\delta}$ to the leading order.

Consider the function
\[
\omega(\lb)=\ln^2\Phi(\lb),
\]
which, for a sufficiently small $\de$, is analytic inside $U_\de$ and maps it conformally onto
a neighborhood of zero.
It has the following expansion at $1$:
\be\la{omexp}
\omega(\lb)=2u\sin^2\frac{\varphi}2\left\{1
-{2\over 3}\left[\cos\varphi+{5\over 4}\right]u+O(u^2)\right\},\quad u=\lb-1,\quad
\lb\in U_\de,
\ee
and
\[
\sqrt{\om}=\sqrt{2}u^{1/2}\sin\frac\varphi 2 (1+O(u)),\quad u=\lb-1.
\]
This expansion is uniform in $\varphi$.

Consider the following mapping of $U_\de$:
\be
\zeta=n^2\om(\lb),\qquad \lb\in U_\de.
\ee

For our analysis below, we need $|\zeta|$ to be uniformly large in $\varphi$ and $\lb$
on the boundary $\partial U_\de$. We see from (\ref{omexp}) that this is indeed so
if $\varphi$ satisfies the condition: ${2s\over n}<\varphi<\pi$, $n>s$, $s>s_0$, with
$s_0$ sufficiently large.

The local parametrix in $U_\de$ is given by the following expression (cf. \cite{DIZ,KVA,K}):
\be
\label{P}
P(\lb)=E(\lb)Q(n^2\omega(\lb))e^{-n\sqrt{\om(\lb)}\sigma_3}f(\lb)^{-\sigma_3/2},
\qquad \lb\in U_\delta,
\ee
where
\be\la{E}
E(\lb)=\frac 1{\sqrt{2}}N(\lb)f(\lb)^{\sigma_3/2}\pmatrix{
                1 & -i\cr
                -i& 1}(\pi n\sqrt{\omega(\lb)})^{\sigma_3/2},
\ee
and the function $Q(\zeta)$ is expressed in terms of modified Bessel and Hankel functions:

1) in the intersection of region $I$ in the $\lb$ plane and $U_\de$
\be
    Q(\zeta) ={1\over 2}
    \pmatrix{
        H_{0}^{(1)}(e^{-i\pi/2}\zeta^{1/2}) &
        H_{0}^{(2)}(e^{-i\pi/2}\zeta^{1/2}) \cr
        \pi \zeta^{1/2} \left(H_{0}^{(1)}\right)'(e^{-i\pi/2}\zeta^{1/2}) &
        \pi \zeta^{1/2} \left(H_{0}^{(2)}\right)'(e^{-i\pi/2}\zeta^{1/2})
    },
\ee

2) region $II$ and $U_\de$
\be
    Q(\zeta) ={1\over 2}
    \pmatrix{
        H_{0}^{(2)}(e^{i\pi/2}\zeta^{1/2}) &
        -H_{0}^{(1)}(e^{i\pi/2}\zeta^{1/2}) \cr
        -\pi \zeta^{1/2} \left(H_{0}^{(2)}\right)'(e^{i\pi/2}\zeta^{1/2}) &
        \pi \zeta^{1/2} \left(H_{0}^{(1)}\right)'(e^{i\pi/2}\zeta^{1/2})
    },
\ee

3) region $III$ and $U_\de$
\be\la{Q}
    Q(\zeta) =
    \pmatrix{
     I_{0} (\zeta^{1/2}) & \frac{i}{\pi} K_{0}(
        \zeta^{1/2})\cr
     \pi i \zeta^{1/2} I_{0}'(\zeta^{1/2}) &
     -\zeta^{1/2} K_{0}'(\zeta^{1/2})},
\ee
where $-\pi<\arg(\zeta)<\pi$.

Using the asymptotic expansions of Bessel and Hankel functions for a large argument, one obtains
uniformly on $\partial U_\delta$:
\be
 \label{delta}
\eqalign{
P(\lb)N(\lb)^{-1}=I+N(\lb)f(\lb)^{\sigma_3/2}
 \left\{\frac
1{8n\sqrt{\omega(\lb)}}\pmatrix{
                -1 & -2i \cr
               -2i & 1}-\frac 3{2^7n^2\omega(\lb)}\pmatrix{
                1 & -4i \cr
               4i & 1}
\right.
\\
\left.
               +O\left(\left[n\sin\frac\varphi 2\right]^{-3}\right)\right\}f(\lb)^{-\sigma_3/2}
               N^{-1}(\lb)=I+\Delta_1+\Delta_2+O\left(\left[
 n\sin\frac\varphi 2\right]^{-3}\right),\qquad \lb\in \partial U_\delta,}
\ee
where $\Delta_1$ and $\Delta_2$ denote the terms with $n\sqrt{\omega(z)}$ and $n^2\omega(z)$, respectively.
(Note that $\Delta_1(\lb)$ and $\Delta_2(\lb)$ are analytic functions in 
$U_\de\setminus\{ 1\}$ with poles
of order 1 at $\lb=1$.)
This is an expansion in the inverse powers of $n\sin\frac\varphi 2$, and
it holds uniformly for  ${2s\over n}<\varphi<\pi$, $n>s$, $s>s_0$, and for $\lb$ on the boundary
$\partial U_\delta$.

Similarly, we define a conformal mapping for the neighborhood $\wt U_\de$:
\[
\omega(\lb)=\ln^2(-\Phi(\lb)),
\]
so that we have
\be\la{omexp2}
\omega(\lb)=-2u\sin^2\frac{\varphi}2\left\{1
+{2\over 3}\left[\cos\varphi+{5\over 4}\right]u+O(u^2)\right\},\quad u=\lb+1,\quad
\lb\in \wt U_\de,
\ee
and
\[
\sqrt{\om}=-i\sqrt{2}u^{1/2}\sin\frac\varphi 2 (1+O(u)),\quad u=\lb+1.
\]

We have for the parametrix in $\wt U_\de$:
\[
\wt{P}(\lb)=\wt{E}(\lb)\sigma_3Q(n^2\omega(\lb))\sigma_3
e^{-n\sqrt{\om(\lb)}\sigma_3}f(\lb)^{-\sigma_3/2}, \quad \lb\in \wt U_\delta,
\]
with
\[
\wt{E}(\lb)=\frac 1{\sqrt{2}}N(\lb)f(\lb)^{\sigma_3/2}\pmatrix{
                1 & i\cr
                i& 1}(\pi n\sqrt{\omega(\lb)})^{\sigma_3/2}.
\]

Therefore,
\be\label{delta1}
\eqalign{
\wt{P}(\lb)N(\lb)^{-1}=I+N(\lb)f(\lb)^{\sigma_3/2}
 \left\{\frac
1{8n\sqrt{\omega(\lb)}}\pmatrix{
                -1 & 2i \cr
               2i & 1}-\frac 3{2^7n^2\omega(\lb)}\pmatrix{
                1 & 4i \cr
               -4i & 1}
\right.
\\
\left.
        +O\left(\left[n\sin\frac\varphi 2\right]^{-3}\right)\right\}f(\lb)^{-\sigma_3/2}
    N^{-1}(\lb)=I+\Delta_1+\Delta_2+O\left(\left[n\sin\frac\varphi 2\right]^{-3}\right),
\qquad \lb\in \partial \wt{U}_\delta,}
\ee
uniformly in $\lb$ and in $\varphi$ for ${2s\over n}<\varphi<\pi$, $n>s$, $s>s_0$.
Similarly to the situation for $U_\de$,
$\Delta_1(\lb)$, $\Delta_2(\lb)$ in $\wt U_\de\setminus\{-1\}$ are analytic functions with poles
of order 1 at $\lb=-1$.

\subsection{Final transformation}

Let
\be\eqalign{
R(\lb)=S(\lb)N^{-1}(\lb),\qquad
\lb\in\bbc\setminus(\overline{U_\de\cup\wt U_\de}\cup\Si_{1,2,3}),\\
R(\lb)=S(\lb)P^{-1}(\lb),\qquad
\lb\in U_\de\setminus\Si_{1,2,3},\\
R(\lb)=S(\lb)\wt P^{-1}(\lb),\qquad
\lb\in \wt U_\de\setminus\Si_{1,2,3}.}\la{Rdef}
\ee
Furthermore, set 
\be\la{Rtilde}
\wt R\equiv {\cal D}_\infty^{-\si_3}R{\cal D}_\infty^{\si_3}.
\ee

It is easy to see that this function has jumps only on
$\partial U_\de$, $\partial\wt U_\de$, and parts of $\Si_1$, and
$\Si_3$ lying outside of the neighborhoods $U_\de$, $\wt U_\de$ (we
denote these parts $\Si_{1,3}^\mathrm{out}$).
Namely,
\be\eqalign{
\wt R_+(\lb)=\wt R_-(\lb){\cal D}_\infty^{-\si_3}
N(\lb)\pmatrix{1&0\cr f(\lb)^{-1}\Phi(\lb)^{-2n}&1}
N(\lb)^{-1}{\cal D}_\infty^{\si_3},\qquad \lb\in\Si_{1,3}^\mathrm{out},\\
\wt R_+(\lb)=\wt R_-(\lb){\cal D}_\infty^{-\si_3}
P(\lb)N(\lb)^{-1}{\cal D}_\infty^{\si_3},\qquad \lb\in \partial U_\de,\\
\wt R_+(\lb)=\wt R_-(\lb){\cal D}_\infty^{-\si_3} \wt P(\lb)N(\lb)^{-1}{\cal D}_\infty^{\si_3},
\qquad \lb\in \partial\wt U_\de.}\la{Rj}
\ee
The jump matrices for $\wt R$ on $\partial U_\de$ and $\partial\wt U_\de$ have the form
\[
I+O(\rho^{-1}),\qquad \rho=n\sin\frac\varphi 2,
\]
uniformly in $\lb$ (as well as in $\al$, $\bt$ in compact sets) and in $\varphi$
provided
\be\la{range}
{2s\over n}<\varphi<\pi,\qquad n>s,\quad s>s_0.
\ee
(A more detailed expansion is given by (\ref{delta}) and (\ref{delta1}).)

Let us now estimate the jump matrix on $\Si_{1,3}^\mathrm{out}$ for $\varphi$
in the range (\ref{range}) with $s_0$ sufficiently large.
Below $\ep'$
will stand for various positive constants independent of $\varphi$, $n$, and $\lb$.
Denote $x=\Re\lb$, $y=\Im\lb$, i.e., $\lb=x+iy$. Choose $\Si_1^\mathrm{out}$ so that $y$ is small,
however, $y>\de'\sin\frac\varphi 2$ with some fixed $\de'>0$.
Multiplying (\ref{Phi}) with its complex conjugate and expanding in $y$ gives:
\be\la{expPhi}
|\Phi(\lb)|^2=1+\frac{2y\sin\frac\varphi 2+O(y^2)}{\sqrt{1-x^2}(\sin^2\frac\varphi 2+
x^2\cos^2\frac\varphi 2)}\qquad \lb\in \Si_1^\mathrm{out}.
\ee
Let $|x|>\ep$ for some $\ep>0$. Then we immediately obtain from (\ref{expPhi}):
\[
|\Phi(\lb)|^2>1+\ep'\sin\frac\varphi 2,
\]
for $\lb\in\Si_1^\mathrm{out}$, and $\varphi$ in the range (\ref{range}).
Now let $|x|\le\ep$. We parametrise $x$ as follows $x=r\sin\frac\varphi 2$. Therefore
$0<r\le \ep/\sin\frac\varphi 2$. If $r>1$, we have
\[
|\Phi(\lb)|^2>1+\frac{r\ep'}{1+r^2}>1+\frac{\ep'}{r}>1+\ep'\sin\frac\varphi 2.
\]
On the other hand, if $r\le 1$, we have, recalling our condition  $y>\de'\sin\frac\varphi 2$,
\[
|\Phi(\lb)|^2>1+\frac{\de'\ep'}{1+r^2}.
\]
Thus we conclude that the estimate $|\Phi(\lb)|^2>1+\ep'\sin\frac\varphi 2$ holds uniformly
for $\lb\in\Si_1^\mathrm{out}$, and $\varphi$ in the range (\ref{range}).
A similar estimate holds on $\Si_3^\mathrm{out}$.
As $N$ and $f$ are bounded on  $\lb\in\Si_{1,3}^\mathrm{out}$ and, in particular, at $\lb=0$,
these estimates immediately imply that the jump matrix on $\Si_{1,3}^\mathrm{out}$
can be written as
\be
I+O(e^{-\ep'\rho}),\qquad \rho=n\sin\frac\varphi 2,
\ee
uniformly for $\lb\in\Si_{1,3}^\mathrm{out}$, and $\varphi$ in the range (\ref{range}).

The above estimates for the jump matrices and the standard analysis of the R-RH problem
(see \cite{DIKairy}) imply that the R-RH problem 
is solvable for large $s_0$ and the solution has the form of the series:
\be
 \label{R}
\wt R(\lb)=I+\sum\limits_{j=1}^{k-1}\wt R_j(\lb)+O(\rho^{-k}),\qquad \wt R_j(\lb)=O(\rho^{-j}),
\ee
uniformly for $\varphi$ in the range (\ref{range}), for $\al$ in a compact subset of the 
$\al$-half-plane $\Re\al>-1/2$, for $\bt$ in a compact subset of 
the $\bt$-plane, and for all $\lb$.

Explicit expressions for $\wt R_k$ are obtained by collecting terms of the same order
in the jump relations. Thus
by (\ref{delta}) we have, in particular, that $\wt R_1$, $\wt R_2$ satisfy the following RH problems.
The functions $\wt R_1$, $\wt R_2$ are analytic in
$\mathbb C\setminus(\partial U_\delta\cup\partial\wt U_\delta)$;
\be\eqalign{
\wt R_{1,+}(\lb)-\wt R_{1,-}(\lb)={\cal D}_\infty^{-\si_3}\Delta_1(\lb){\cal D}_\infty^{\si_3},\\
\wt R_{2,+}(\lb)-\wt R_{2,-}(\lb)=\wt R_{1,-}(\lb){\cal D}_\infty^{-\si_3}
\Delta_1(\lb){\cal D}_\infty^{\si_3}+{\cal D}_\infty^{-\si_3}\Delta_2(\lb){\cal D}_\infty^{\si_3},
\quad
\lb\in\partial U_\delta\cup\partial\wt U_\delta;}
\ee
and $\wt R_1\to 0$, $\wt R_2\to 0$ as $\lb\to\infty$.

As $\Delta_1(\lb)$ is analytic in neighborhoods of the endpoints
except for the simple poles at $\pm 1$, we have an expansion
\be\eqalign{
{\cal D}_\infty^{-\si_3}\Delta_1(\lb){\cal D}_\infty^{\si_3}
=\frac{A^{(1)}}{\lb-1}+A^{(2)}+A^{(3)}(\lb-1)+O((\lb-1)^2),\quad \mbox{ as } \lb\to1;\\
{\cal D}_\infty^{-\si_3}\Delta_1(\lb){\cal D}_\infty^{\si_3}
=\frac{B^{(1)}}{\lb+1}+B^{(2)}+B^{(3)}(\lb+1)+O((\lb+1)^2), \quad \mbox{ as } \lb\to -1,}
\ee
where $A$ and $B$ are constant matrices.
It is now easy to verify that the RH problem for $\wt R_1$ has the following solution:
\be
\label{r1}
\wt R_1(\lb)=
\cases{\frac{A^{(1)}}{\lb-1}+\frac{B^{(1)}}{\lb+1},&
for
$\lb\in\bbc\setminus\overline{U_\delta}\cup\overline{\wt U_\delta}$ ,\cr
\frac{A^{(1)}}{\lb-1}+\frac{B^{(1)}}{\lb+1}-{\cal D}_\infty^{-\si_3}
\Delta_1(\lb){\cal D}_\infty^{\si_3},
& for $\lb\in U_\delta \cup\wt U_\delta$ . }
\ee

For the function $\Delta_2$ we can write similarly:
\be\eqalign{
{\cal D}_\infty^{-\si_3}\Delta_2(\lb){\cal D}_\infty^{\si_3}
=\frac{C^{(1)}}{\lb-1}+C^{(2)}+O(\lb-1),\quad \mbox{ as } \lb\to1;\\
{\cal D}_\infty^{-\si_3}\Delta_2(\lb){\cal D}_\infty^{\si_3}
=\frac{D^{(1)}}{\lb+1}+D^{(2)}+O(\lb+1), \quad \mbox{ as } \lb\to -1,}
\ee
where $C$ and $D$ are constant matrices.
A similar expression to (\ref{r1}) can now be written for $R_2$. However, we will need it
only in the limit $\lb\to\pm 1$. We then obtain:
\be\la{R2pm1}\eqalign{
\wt R_2(1)=-{1\over 4}[A^{(1)},B^{(1)}]-{1\over 2}(B^{(2)}B^{(1)}+B^{(1)}A^{(2)}-D^{(1)})+
{A^{(2)}}^2+A^{(3)}A^{(1)}-C^{(2)}\\
\wt R_2(-1)={1\over 4}[A^{(1)},B^{(1)}]+{1\over 2}(A^{(2)}A^{(1)}+A^{(1)}B^{(2)}-C^{(1)})+
{B^{(2)}}^2+B^{(3)}B^{(1)}-D^{(2)},}
\ee
where $[A,B]=AB-BA$.
We also obtain from (\ref{r1}),
\be\la{R1pm1}
\wt R_1(1)={1\over 2}B^{(1)}-A^{(2)},\qquad \wt R_1(-1)=-{1\over 2}A^{(1)}-B^{(2)}.
\ee

\subsection{Asymptotic form of the differential identity (\ref{di3})}
We will now compute asymptotics of the r.h.s. of (\ref{di3}).
First, we determine the components of the matrices $A^{(j)},\dots,D^{(j)}$ we need for the calculation
below.
First, the expansions for $\om(\lb)$ in Section \ref{local} can be written as
\[\eqalign{
{1\over\sqrt{\om(\lb)}}={1\over (2u)^{1/2}\sin\frac\varphi 2}(1+\om_{1+}u+
\om_{2+}u^2+O(u^3)),\qquad u=\lb-1,\\
{1\over\sqrt{\om(\lb)}}={1\over -i(2u)^{1/2}\sin\frac\varphi 2}(1+\om_{1-}u+
\om_{2-}u^2+O(u^3)),\qquad u=\lb+1,}
\]
with
\[
\om_{1\pm}=\pm{1\over 3}\left(\cos\varphi+{5\over 4}\right).
\]
(The explicit expression for $\om_{2\pm}$ will not be needed below.)
Therefore, expanding $N(\lb)$ we obtain from (\ref{delta})
\be\eqalign{
A^{(1)}={1\over 16\rho}\pmatrix{1& -i\cr -i&-1},\qquad
A^{(2)}={1\over 16\rho}\left[\om_{1+}\pmatrix{1& -i\cr -i&-1}+\Gamma_0\right],\\
\Gamma_0={1\over 2}\pmatrix{\gamma_0& -i(6+\gamma_0-8{\cal D}_{1+})\cr
-i(6+\gamma_0+8{\cal D}_{1+})& -\gamma_0}}
\ee
with
\[
\rho= n \sin\frac\varphi 2,\qquad \gamma_0=-{5\over 2}+4{\cal D}_{1+}^2.
\]
Here ${\cal D}_{1+}$ is given by (\ref{D1}).
Furthermore,
\be\eqalign{
A^{(3)}=
{1\over 16\rho}\left[\om_{2+}\pmatrix{1& -i\cr -i&-1}+\om_{1+}\Gamma_0
+\Gamma_1\right],\\
\Gamma_1={1\over 4}\pmatrix{\gamma_1& -i(\gamma_2-8{\cal D}_{2+}
-{16\over 3}{\cal D}_{1+}^3)\cr
-i(\gamma_2+8{\cal D}_{2+}
+{16\over 3}{\cal D}_{1+}^3)& -\gamma_1},}
\ee
with
\[
\gamma_1={11\over 8}-2{\cal D}_{1+}^2+8{\cal D}_{1+}{\cal D}_{2+}+
{4\over 3}{\cal D}_{1+}^4,\qquad \gamma_2=\gamma_1+8{\cal D}_{1+}^2-3.
\]
We obtain the matrices $B^{(j)}$, $j=1,2,3$, by taking the matrices
$A^{(j)}$, $j=1,2,3$, and replacing $\rho$ with $-\rho$; $i$ with $-i$;
$\om_{1,2+}$ with $\om_{1,2-}$; ${\cal D}_{1,2+}$ with ${\cal D}_{1,2-}$;
and the prefactor ${1\over 2}$ of $\Gamma_0$ with $-{1\over 2}$.

The components of $C$ are also obtained from (\ref{delta}).
For $C$ (and similarly for $D$) we will only need the following combination
of matrix elements:
\be
C^{(1)}_{21}-C^{(1)}_{12}=\frac{-3i}{2^5\rho^2},\qquad
C^{(2)}_{21}-C^{(2)}_{12}=\frac{-3i}{2^5\rho^2}(2\om_{1+}+{\cal D}_{1+}^2).
\ee
Similarly,
\be
D^{(1)}_{21}-D^{(1)}_{12}=\frac{-3i}{2^5\rho^2},\qquad
D^{(2)}_{21}-D^{(2)}_{12}=\frac{3i}{2^5\rho^2}(-2\om_{1-}+{\cal D}_{1-}^2).
\ee

We now need to evaluate the asymptotics of $\tr C_+C_-$ and $\tr C_1(C_++C_-)$
from (\ref{di3}).
Tracing back the transformations of the RH problem, we see that $Y(z)$ for $z$ close to $z_+$,
so that $\lb$ is close to $1$ in the region $III$,
is given by
\[
Y(z)=\ga^{n\si_3}T_0 R P \Psi^{n\si_3},\qquad
P(\lb)=E(\lb)Q(n^2\omega(\lb))e^{-n\sqrt{\om(\lb)}\sigma_3}f(\lb)^{-\sigma_3/2},
\]
where $E$ is given by (\ref{E}), and $Q$ by (\ref{Q}).
Substituting this into (\ref{Cpm}) and expanding Bessel functions at $\ze=0$, we obtain:
\be\la{C+}
C_+=\frac{\rho}{2i}\ga^{n\si_3}T_0{\cal D}_\infty^{\si_3}
\wt R(1)
\pmatrix{i & 1\cr 1 & -i}
\wt R(1)^{-1}
{\cal D}_\infty^{-\si_3}T_0^{-1}\ga^{-n\si_3}.
\ee
Here we can write
\[
\wt R(1)=I+\wt R_1(1)+\wt R_2(1)+O(\rho^{-3}),\qquad
\wt R(1)^{-1}=I-\wt R_1(1)-\wt R_2(1)+\wt R_1^2(1)+O(\rho^{-3}).
\]
For definitiveness, we assume that the roots in (\ref{N}) are chosen with the arguments
from $0$ to $2\pi$, and that the point $\lb$ corresponding to the limit in (\ref{Cpm})
approaches $\lb=1$ from above.

For $C_-$ we similarly have
\be\la{C-}
C_-=\frac{\rho}{-2i}\ga^{n\si_3}T_0{\cal D}_\infty^{\si_3}
\wt R(-1)
\pmatrix{-i & 1\cr 1 & i}
\wt R(-1)^{-1}
{\cal D}_\infty^{-\si_3}T_0^{-1}\ga^{-n\si_3}.
\ee
Therefore, we now easily obtain
\be
\tr C_+C_-=
\rho^2\left[1+t_1+t_2+O\left({1\over\rho^3}\right)\right],
\ee
where (recall first (\ref{ABC}), and then (\ref{R1pm1}) and the above expressions
for $A$, $B$)
\be\la{t1}
t_1=i(\wt R_{1,\,21}(1)-\wt R_{1,\,12}(1)-
[\wt R_{1,\,21}(-1)-\wt R_{1,\,12}(-1)])=
-{2\al\over \rho}\left(1-\sin\frac\varphi 2\right).
\ee
The expression for $t_2$ needs more work. First, we obtain it in the form
\be\la{t2}
t_2=i(\wt R_{2,\,21}(1)-\wt R_{2,\,12}(1)-
[\wt R_{2,\,21}(-1)-\wt R_{2,\,12}(-1)])+\Om,
\ee
where
\[\eqalign{
4\Om=\tr\left\{\wt R_1(1)\left[ \wt R_1(1),\pmatrix{-i&1\cr1&i}\right]\pmatrix{i&1\cr1&-i}\right\}\\
+
\tr\left\{\wt R_1(-1)\left[ \wt R_1(-1),\pmatrix{i&1\cr1&-i}\right]\pmatrix{-i&1\cr1&i}\right\}+
\tr\left\{\left[ \wt R_1(1),\pmatrix{i&1\cr1&-i}\right]\left[\wt R_1(-1),\pmatrix{-i&1\cr1&i}\right]\right\}.}
\]
Denoting the matrix elements
\[
\wt R_1(1)=\pmatrix{a& ib\cr ic& -a},\qquad \wt R_1(-1)=\pmatrix{\hat a & -i\hat b\cr -i\hat c & -\hat a}
\]
we obtain after a simple algebraic computation that
\[
\Om={1\over 4}(b-c+\hat b-\hat c)^2+(a-\hat a)^2-
(c+\hat c)(b+\hat b)-a(\hat b+\hat c)-\hat a(b+c).
\]
Using again (\ref{R1pm1}) and the above expressions
for $A$, $B$, we deduce from this formula that
\be\la{Tau}
\Om=\frac{-1}{2^7 \rho^2}(4({\cal D}_{1+}^2+{\cal D}_{1-}^2)+
2(\om_{1+}-\om_{1-})+5-16({\cal D}_{1+}+{\cal D}_{1-})^2).
\ee

On the other hand, we obtain from (\ref{R2pm1}) and the above expressions
for $A$, $B$, $C$, $D$:
\be\la{R2t2}
i(\wt R_{2,\,21}(1)-\wt R_{2,\,12}(1)-
[\wt R_{2,\,21}(-1)-\wt R_{2,\,12}(-1)])=
\frac{1}{2^7 \rho^2}(9-22(\om_{1+}-\om_{1-})-12({\cal D}_{1+}^2+{\cal D}_{1-}^2)).
\ee
Thus, substituting (\ref{Tau}), (\ref{R2t2}) into (\ref{t2}) and then using the expressions for
$\om_{1\pm}$, ${\cal D}_{1+}^2$, we obtain
\be\la{t2final}
t_2=
\frac{1}{2^5\rho^2}(1-6(\om_{1+}-\om_{1-})+8{\cal D}_{1+}{\cal D}_{1-})=
\frac{1}{\rho^2}\left(
\al^2\left(1-\sin\frac\varphi 2\right)^2+\bt^2\cos^2\frac\varphi 2-{1\over 4}\cos^2\frac\varphi 2\right).
\ee

Collecting together (\ref{t1}), (\ref{t2final}), we finally have
\be\la{CpCmas}
\tr C_+C_-=n^2 \sin^2\frac\varphi 2-2\al n \sin\frac\varphi 2 \left(1-\sin\frac\varphi 2\right)
-{1\over 4}\cos^2\frac\varphi 2+\al^2\left(1-\sin\frac\varphi 2\right)^2+\bt^2\cos^2\frac\varphi 2+
O(\rho^{-1}).
\ee

We now turn our attention to the quantity $\tr C_1(C_+ +C_-)$.
Recall that $z=1$ corresponds to $\lb=\infty$. We, therefore, immediately obtain
\be
C_1=\al Y(1)\si_3 Y(1)^{-1}=\al\ga^{n\si_3}T_0\si_3 T_0^{-1}\ga^{-n\si_3}.
\ee
In $C_+ +C_-$ we now need to take into account only the terms of order no less than $\wt R_1$.
Using (\ref{C+}), (\ref{C-}), and at the last step (\ref{t1}), we obtain
\be
\tr C_1(C_+ + C_-)=\al n\sin\frac\varphi 2 (2+t_1)+O(\rho^{-1})=
2\al n\sin\frac\varphi 2 -2\al^2\left(1-\sin\frac\varphi 2\right)+O(\rho^{-1}).
\ee
Substituting this expression and (\ref{CpCmas}) into (\ref{di3}), we finally have

\noindent
{\bf Proposition 9} {\it As $n\to\infty$,
\be\la{di3as}
\frac{d^2}{d\varphi^2}\ln D_n(\varphi)=
-\frac{n^2}{4\cos^2\frac\varphi 2}-(\al n+\al^2)\frac{1-\sin\frac\varphi 2}{2\cos^2\frac\varphi 2}
+\frac{1+4(\al^2-\bt^2)}{16\sin^2\frac\varphi 2}+O\left({1\over n \sin\frac\varphi 2}\right),
\ee
uniformly for ${2s\over n}<\varphi<\pi$, $n>s$, $s>s_0$, for some $s_0>0$, and
uniformly in compact subsets of the half-plane $\Re\al>-1/2$ and 
of the plane $\bt\in\bbc$.
}

\section{Proof of Theorem 1}
The asymptotic evaluation of the Toeplitz determinant $D_{n}(\varphi)$
is based on the integration of the differential identity
(\ref{di3as}) from $\varphi$ to $\pi-\ep$ with a small positive $\ep$.
We have:
\be\la{Dinteg}
(\pi-\ep-\varphi)(\ln D_n)'|_{\pi-\ep}- \ln D_n(\pi-\ep) +\ln D_n(\varphi)=
\int_\varphi^{\pi-\ep}d\theta \int_\theta^{\pi-\ep} I(\phi)d\phi,
\ee
where $I(\varphi)$ is the r.h.s. of (\ref{di3as}).
Fix $n$. Calculating the integral on the r.h.s. of (\ref{Dinteg}), substituting for
$\ln D_n(\pi-\ep)$ the expansion (\ref{Dnpi}) (and for $(\ln D_n)'|_{\pi-\ep}$ its derivative),
and taking the limit $\ep\to 0$, we obtain
\be
\label{Dnn}\eqalign{
\ln D_n(\varphi)=n^2\ln\cos\frac\varphi 2+ 2(\al n + \al^2)\ln\left(1+\sin\frac\varphi 2\right)
-\frac 14\ln n+\frac 1{12}\ln 2+3\zeta'(-1)\\
-2\alpha^2\ln 2-\left(\frac 14-\beta^2+\alpha^2\right)\ln\sin\frac\varphi 2+
O\left(\frac 1{n\sin\frac{\varphi}2}\right)+\delta_n.}
\ee
uniformly for $\frac {2s}n<\varphi<\pi$, $n>s$, $s>s_0$, and where $\de_n\to 0$ as $n\to\infty$.
This expansion is uniform in compact subsets of the $\al$-half-plane $\Re\al>-1/2$ and 
of the $\bt$-plane.
Now substituting (\ref{Cn}) and (\ref{Dnn}) into (\ref{scalinglimit}) and taking the limit $n\to\infty$,
we obtain (\ref{Ps}). $\Box$

\section{Bessel kernel. Proof of Theorem 4.}
In this section we set $\bt=0$. The weight (\ref{weight}) is then
\be
\label{newweight0}
f(z,\varphi)=|z-1|^{2\alpha}, \quad z=e^{i\theta}, \quad \varphi\le\theta\le 2\pi-\varphi,
\ee
and $f(z,\varphi)=0$ on the rest of the unit circle. Note that $f$ is an even function of the angle
$\theta$. Let
\be
\label{omega}
\omega(x,\varphi)=\frac{f(e^{i\theta},\varphi)}{|\sin\theta|}=
2^{\al}\frac{(1-x)^{\al}}{\sqrt{1-x^2}}, \quad x=\cos\theta.
\ee
This function is supported on $[-1,\cos\varphi]$. Consider the Hankel determinant
with symbol $\omega(x,\varphi)$:
\be\la{Hankel}
\eqalign{
D_n^H(\varphi)=\det\left(\int\limits_{-1}^{\cos\varphi} x^{j+k}\omega(x,\varphi)dx\right)_{j,k=0}^{n-1}
\\
={1\over n!}\int_{-1}^{\cos\varphi}\cdots\int_{-1}^{\cos\varphi}
\prod_{1\le j<k\le n}(x_j-x_k)^2\prod_{j=1}^n \omega(x_j,\varphi)dx_j.}
\ee
There holds the following

\noindent
{\bf Lemma 10}
{\it 
Let $K_{Bessel2}^{(a)}$ be the operator acting on $L^2(0,(2s)^2)$, with kernel (\ref{Bessel2}). Then
\be
\label{scaling2}
\det(I-K_{Bessel2}^{(\al-1/2)})_{L^2(0,(2s)^2)}=\lim_{n\to \infty}\frac{D_n^H(\frac{2s}{n})}{D_n^H(0)}.
\ee
}

\noindent {\it Proof.} 
First,
as in the proof of the formula (\ref{Dform}) of Lemma 6, one obtains that
\be\la{detH1}
D_n^H(\varphi)=D_n^H(0)\det(I-\wt K_n)_{L^2(\cos\varphi,1)}
\ee
where $\wt K_n$ is the operator acting on $L^2((\cos\varphi,1),dx)$ with kernel
\be\la{newKn}
\wt K_n(x,y)=\sqrt{\omega(x)\omega(y)}\frac{\varkappa_{n-1}}{\varkappa_n}
\frac{P_n(x)P_{n-1}(y)-P_{n-1}(x)P_{n}(y)}{x-y},
\ee
where $P_k(x)=\varkappa_k x^k+\cdots$, $k=N_0,N_0+1,\dots$, with some $N_0\ge 0$,
are the polynomials
orthonormal on $[-1,1]$ w.r.t. $w(x,0)$:
\[
\int_{-1}^{1}P_k(x)x^m\omega(x,0)dx=\varkappa_k^{-1}\delta_{km}, \quad m=0,1,\dots,k.
\]

The choice of the function (\ref{omega}) implies the following Szeg\H o relations
between $P_k(x)$ and the polynomials $q_k(z)=\chi_k z^k+\cdots$,
given by (\ref{q0}) (see Lemma 2.5. of \cite{DIKT}):
\be
\label{Pn}
P_k(x)=\frac{1}{\sqrt{2\pi(1+q_{2k}(0)/\chi_{2k})}}(z^{-k}q_{2k}(z)+z^k q_{2k}(z^{-1})),
\ee
and for the leading coefficients,
\[
\varkappa_k=2^k\chi_{2k}\sqrt{\frac{1+q_{2k}(0)/\chi_{2k}}{2\pi}}.
\]
Note that $\widehat q_n(z)=q_n(z)$ as in our case $f(e^{i\theta})=f(e^{-i\theta})$.

Similarly to the proof of Lemma 6, we now set 
$x=\cos(2u/n)$, $y=\cos(2v/n)$, fix $0<u<s$ and $0<v<s$ and
consider the limit of $\wt K_n(x,y)$ as $n\to\infty$. We will now show that this double-scaling
limit gives the kernel of $K_{Bessel2}^{(\al-1/2)}$.

First, it follows from Theorem 1.8. of \cite{DIKT} that 
\[
\frac{q_{2n}(0)}{\chi_{2n}}=O\left(\frac{1}{n}\right),\qquad 
\frac{\varkappa_{n-1}}{\varkappa_n}=\frac 12+O\left(\frac{1}{n}\right).
\]
Moreover, we have,
\[
x=\cos\left({2u\over n}\right)=1-{2u^2\over n^2}+o(n^{-2}),\qquad z(x)=e^{2iu/n},
\]
and similarly for $y=\cos(2v/n)$.
Therefore, by (\ref{omega}), 
\be\la{omega2}
\sqrt{\om(x,0)\om(y,0)}=\left({2\over n}\right)^{2\al-1}(uv)^{\al-1/2}(1+O(n^{-1})).
\ee
Moreover, taking $z_1=e^{2iu/n}\in\bbc_+$ and using the expression (\ref{qn}) of Lemma 5, we have
\[\eqalign{
q_{2(n+k)}(z_1)=(2n)^\al\frac{\Gamma(1+\al)}{\Gamma(1+2\al)}
\phi(1+\al,1+2\al,4iu(1+k/n))(1+O(1/n))\\
=(2n)^\al\frac{\Gamma(1+\al)}{\Gamma(1+2\al)}
\left[\phi(1+\al,1+2\al,4iu)+\frac{4iuk}{n}\phi'(1+\al,1+2\al,4iu)\right](1+O(1/n))
,\\ k=0,-1.}
\]
Substituting this into (\ref{Pn}), we then obtain an expression for 
$P_n(x)P_{n-1}(y)-P_{n-1}(x)P_{n}(y)$ in terms of the confluent hypergeometric
functions and their derivatives at $\pm 4iu$, $\pm 4iv$. 
Removing the derivatives with the help of the standard relation
\[
\phi'(a,c,x)={a\over x}(\phi(a+1,c,x)-\phi(a,c,x))
\]
reducing then, by Kummer's transformation (\ref{Kummer}),
the terms with the arguments $-4iu$, $-4iv$ to functions
of the arguments $4iu$, $4iv$, and making use of the following standard recurrence relation
\[
(c-a)\phi(a-1,c,x)+(2a-c+x)\phi(a,c,x)-a\phi(a+1,c,x)=0,
\]
we obtain
\be\eqalign{
P_n(x_1)P_{n-1}(x_2)-P_{n-1}(x_1)P_{n}(x_2)=
\frac i{\pi}\frac{\Gamma^2(1+\alpha)}{\Gamma^2(1+2\alpha)}(2n)^{2\alpha}
e^{-2i(u+v)}\frac1 n\left(1+O\left( \frac 1n\right)  \right)\\
\times
\left\{ (u+v)[\phi(1+\alpha,1+2\alpha,4iu)\phi(\alpha,1+2\alpha,4iv)-
\phi(1+\alpha,1+2\alpha,4iv)\phi(\alpha,1+2\alpha,4iu)]
\right.\\
\left.
+(u-v)\left[\phi(1+\alpha,1+2\alpha,4iu)\phi(1+\alpha,1+2\alpha,4iv)-
\phi(\alpha,1+2\alpha,4iu)\phi(\alpha,1+2\alpha,4iv)\right]   \right\}.}
\ee
Thus, in the difference $P_n(x)P_{n-1}(y)-P_{n-1}(x)P_{n}(y)$ the main terms 
in $n$ dropped out leaving the ones of order $1/n$. 
Moreover, we did not need to know expressions for the terms 
$O(1/n)$ in Lemma 5, as their contribution to the terms of order $1/n$ dropped out as well.

We now employ the recurrence relations (\ref{recurrence})
to express $\phi(\al,1+2\al,x)$ and $\phi(1+\alpha,1+2\alpha,x)$ in terms of
$\phi(\al,2\al,x)$ and $\phi(1+\alpha,2+2\alpha,x)$ and then use the connection with Bessel
functions (\ref{confbessel}). We obtain recalling (\ref{omega2}):
\[
\wt K_n(x,y)=\frac {n^2}2\frac{[uJ_{\alpha+1/2}(2u)J_{\alpha-1/2}(2v)-
vJ_{\alpha+1/2}(2v)J_{\alpha-1/2}(2u)]}{u^2-v^2}\left(1+O\left( \frac 1n\right)  \right),
\]
which leads to
\be
\label{kernelB}
\lim\limits_{n\to \infty}\left(-\frac{1}{2n^2}\right)\wt K_n(x,y)=- 
K_{Bessel2}^{(\al-1/2)}((2u)^2,(2v)^2),
\ee
where $-K_{Besssel2}^{(\al-1/2)}$ acts on $((2s)^2,0)$ (note the reversed direction) and
\be\la{Bessel22}
K_{Bessel2}^{(a)}(x,y)=
\frac{\sqrt{x}J_{a+1}(\sqrt{x})J_a(\sqrt{y})-
\sqrt{y}J_{a+1}(\sqrt{y})J_{a}(\sqrt{x})}{2(x-y)},\quad a=\al-1/2,
\ee
which, by the relation $zJ_{a+1}(z)=aJ_a(z)-zJ'_a(z)$, is equivalent to (\ref{Bessel2}).

The convergence of the determinants follows from the convergence of the kernels
as in Lemma 6, and we obtain the statement (\ref{scaling2}) from 
(\ref{detH1}). $\Box$

We now evaluate the r.h.s. of (\ref{scaling2}). First, note that $D_n^H(0)$ is a Hankel determinant
whose symbol $\om(x,0)$ (\ref{omega}) is supported on $[-1,1]$ and has two Fisher Hartwig singularities
at $x=-1$ and $x=1$. Therefore, the asymptotics of $D_n^H(0)$ are given by a particular case
of Theorem 1.20 from \cite{DIKT}. Namely,
\be
\label{DH0}
D_n^H(0)=\frac{\pi^{n+\alpha/2}G(1/2)}{G(1/2+\alpha)}
2^{-(n-1)^2-\frac{\alpha^2}{2}+\frac{3\alpha}{2}}n^{\frac{\alpha^2-\alpha}{2}}\left(1+o(1)\right),\qquad
n\to\infty,
\ee
uniformly in compact subsets of the half-plane $\Re\al>-1/2$.

In order to evaluate  $D_n^H(2s/n)$,  we use a connection formula 
between Hankel and Toeplitz determinants
given by Theorem 2.6 in \cite{DIKT}. The formula
is written in terms of the matrix elements of $Y^{(2n)}(z)$ (\ref{Y}) 
and for $\varphi=2s/n$ as follows:
\be
\label{Dnnn}
\left(D_n^H\left(\frac{2s}n\right)\right)^2=\frac{\pi^{2n}}{2^{2(n-1)^2}}
\frac{(1+Y_{11}^{(2n)}(0))^2}{Y_{11}^{(2n)}(1)Y_{11}^{(2n)}(-1)}D_{2n}\left(\frac{2s}n\right).
\ee

The asymptotic expression for the Toeplitz determinant $D_{2n}\left(\frac{2s}n\right)$
is given 
(uniformly in compact subsets of the half-plane $\Re\al>-1/2$)
by (\ref{Dnn}) with $n$ replaced by $2n$ and with $\varphi$ set to be $2s/n$:
\be\la{DTs}\eqalign{
\ln D_{2n}\left(\frac{2s}n\right)=-2s^2+4\alpha s -\left(\alpha^2+\frac 14\right)\ln s+
\alpha^2\ln n -\left( 2\alpha^2+\frac 14\right)\ln 2\\
+{1\over2}\ln\pi+2\ln G(1/2)+
O\left(\frac 1s \right)+\hat\delta_n(s),\qquad n\to\infty,}
\ee
where $\hat\delta_n(s)\to 0$ as $n\to\infty$.

It now remains to estimate $Y^{(2n)}_{11}(z)$ at $z=-1,0,1$. The $\lb$-images of these points are
$\lb=0,-i\tan\frac{\varphi}{2},\infty$, respectively. All of them lie in the regions where
$Y^{(2n)}(z)$ is approximated by the outside parametrix $N(\lb)$.
From the expressions (\ref{T},\ref{Ttilde},\ref{S1}--\ref{S3},\ref{Rdef},\ref{Rtilde},\ref{R}),
we obtain:
\be\la{Yas}
Y(z)^{(2n)}=\gamma^{2n\si_3}N(i\tan\frac{\varphi}{2})^{-1}\left(I+
{\cal D}_\infty^{\si_3}O(\rho^{-1}){\cal D}_\infty^{-\si_3}\right)
N(\lb(z))
\pmatrix{1& 0\cr b(z) & 1}\Psi(z)^{2n\si_3},
\ee
which is valid for $z$ in neighborhoods of $z=-1,0,1$. In a neighborhood of $-1$ we assume
that $|z|>1$ and then $b(z)=f(z)^{-1}\Psi(z)^{-4n}z^{2n}$. In neighborhoods of $0$ and $1$,
$b(z)=0$.

We will need the values of ${\cal D}(\lb)$ at $\lb=0,-i\tan\frac{\varphi}{2},\infty$.
Analyzing the integral in (\ref{D}) with $\bt=0$, we obtain 
\be\label{DD1}
{\cal D}\left(i\tan\frac \varphi 2\right)=D\left(-i\tan\frac \varphi 2\right)
=\left(1+\sin\frac \varphi 2\right)^\alpha,\qquad {\cal D}(0)=2^\alpha.
\ee
From (\ref{Dinfty}) we have
\be\label{DD3}
{\cal D}_\infty=\left(\frac{4\sin\frac\varphi 2}{1+\sin\frac\varphi 2}\right)^\al.
\ee

Noting that in the definition of $N(i\tan\frac \varphi 2)$,
$a(i\tan\frac{\varphi}{2})=e^{i(\pi-\varphi)/4}$, we can write the following expression
for the $11$ element of (\ref{Yas}):
\be
\label{Y11}\eqalign{
Y_{11}^{(2n)}(z)=\frac 12\gamma^{2n}\Psi(z)^{2n}\left(1+\sin\frac\varphi 2\right)^\alpha
\left[{\cal D}(\lb(z))^{-1}
\left(e^{i(\varphi-\pi)/4}a(\lambda)+e^{-i(\varphi-\pi)/4}a^{-1}(\lambda) \right)\right.\\
\left. 
+b(z){\cal D}(\lb(z))i^{-1}
\left(e^{i(\varphi-\pi)/4}a(\lambda)-e^{-i(\varphi-\pi)/4}a^{-1}(\lambda) \right)
\right](1+O(\rho^{-1})).
}
\ee 
Since by (\ref{psi}),
\[
\Psi(-1)=-1,\qquad \Psi(0)=\ga^{-1},\qquad \Psi(1)=\ga^{-1}\left(1+\sin\frac\varphi 2\right),
\]
we obtain from (\ref{Y11})
\be\la{Y1}
Y_{11}^{(2n)}(-1)=\ga^{2n} 2^{-\al}\left(1+\sin\frac\varphi 2\right)^\al
\left[\cos{\varphi\over 4}+\sin{\varphi\over 4}\right](1+O(\rho^{-1}))=2^{-\al}[1+O(s^{-1})],\quad 
n>s^2,
\ee
where the second equation is obtained by substituting $\varphi=2s/n$,
and considering $n>s^2$ and $s$ large.
Similarly,
\be
Y_{11}^{(2n)}(0)=\sin{\varphi\over 2}(1+O(\rho^{-1}))={s\over n}[1+O(s^{-1})],\quad n>s,
\ee
and
\be\la{Y3}
Y_{11}^{(2n)}(1)= \frac{\left(1+\sin\frac\varphi 2\right)^{2(n+\al)}}
{2^{2\al}\sin^{\al}\frac\varphi 2}\cos{\varphi-\pi\over 4}(1+O(\rho^{-1}))=
\frac{e^{2s}n^\al}{2^{2\al+1/2}s^\al}[1+O(s^{-1})],\quad n>s^2.
\ee
Note that (\ref{Y1}--\ref{Y3}) are uniform in compact subsets of the half-plane $\Re\al>-1/2$.

Substituting (\ref{Y1}--\ref{Y3}) and (\ref{DTs}) into (\ref{Dnnn}) and taking the square root,
we obtain
\be\la{DHs}
D_n^H\left(\frac{2s}n\right)=
\pi^{n+{1\over 4}}n^{\al^2-\al\over 2}
2^{-(n-1)^2+{3\over 2}\al-\al^2+{1\over 8}}G(1/2)
s^{-{1\over 2}\left(\al-{1\over 2}\right)^2}
e^{-s^2+(2\al-1)s}[1+O(s^{-1})],\quad n>s^2.
\ee
The branch of the square root is fixed by the fact that $D_{2n}^H>0$ for $\al\in\bbr$,
and by the uniformity of the asymptotic expansion in $\al$.
Finally, substituting (\ref{DHs}) and (\ref{DH0}) into (\ref{scaling2}),
we finish the proof of Theorem 4. $\Box$

\section{Appendix}
Here we show that the operator $K^{(\al,\bt)}$ with kernel 
(\ref{kernel}) on $L^2(-s,s)$, where $\Re\al>-1/2$ and $\al\pm\bt\neq -1,2,\dots$, is trace class.
Note first that since $g_{\bt}^{1/2}(x)$ is bounded on $(-s,s)$, it is sufficient to
show that the operator $\widehat K$ with the following kernel
\be\la{k-int-1}\eqalign{
\widehat K(x,y)=\frac{|x|^\al |y|^\al}{x-y}
[e^{i(y-x)}\phi(1+\alpha+\beta,1+2\alpha,2ix)\phi(1+\alpha-\beta,1+2\alpha,-2iy)\\
-e^{i(x-y)}\phi(1+\alpha+\beta,1+2\alpha,2iy)\phi(1+\alpha-\beta,1+2\alpha,-2ix)]}
\ee
is trace class. Expanding the confluent hypergeometric functions in series,
\[
e^{-ix}\phi(1+\alpha+\beta,1+2\alpha,2ix)=
\sum_{n=0}^\infty\mu_n x^n,\qquad 
e^{ix}\phi(1+\alpha-\beta,1+2\alpha,-2ix)=
\sum_{m=0}^\infty\lb_m x^m,
\]
where $\mu_n$, $\lb_n$ are determined using (\ref{phidef}), we can write the kernel (\ref{k-int-1})
in the form:
\be
\widehat K(x,y)=|x|^\al |y|^\al
\sum_{n=0}^\infty\sum_{m=0}^\infty\mu_n\lb_m
\frac{x^n y^m -y^n x^m}{x-y}.
\ee
We will now show that the trace norm in $L^2(-s,s)$
\be\la{normineq}
\| |x|^\al \frac{x^n y^m -y^n x^m}{x-y} |y|^\al \|_1\le
C n(m+1)s^{2\Re\al+m+n},\qquad m\ge 0,\quad n\ge 1 
\ee
for some $C>0$.
Together with the straightforward estimates
\[
|\mu_n|, |\lb_n| \le \frac{n^b}{n!},\qquad n\ge 1,
\]
for some $b\in\bbr$, the inequality
(\ref{normineq}) implies that $\widehat K$, and hence $K^{(\al,\bt)}$, is trace class. 

To prove (\ref{normineq}), set first $m=0$, $n\ge 1$. Then we have for some $C>0$
\be\la{normineq1}
\eqalign{
\| |x|^\al \frac{x^n - y^n}{x-y} |y|^\al \|_1\le
\sum_{k=0}^{n-1}\| |y|^{k+\Re\al} \|_{L^2(-s,s)} \| |x|^{n-k-1+\Re\al} \|_{L^2(-s,s)}=\\
\sum_{k=0}^{n-1}\left(\frac{2 s^{2(k+\Re\al)+1}}{2(k+\Re\al)+1}\right)^{1/2}
\left(\frac{2 s^{2(n-k+\Re\al)-1}}{2(n-k+\Re\al)-1}\right)^{1/2}\le\\
C\sum_{k=0}^{n-1}s^{{1\over2}(2(k+\Re\al)+1)+{1\over2}(2(n-k+\Re\al)-1)}=
C n s^{n+2\Re\al},}
\ee
which gives (\ref{normineq}) for $m=0$, $n\ge 1$. 
If $m\ge 1$, $n\ge 1$, we can assume that $n>m$ and write
\[
|x|^\al \frac{x^n y^m -y^n x^m}{x-y} |y|^\al =
 |x|^{\al+m} \frac{x^{n-m} - y^{n-m}}{x-y} |y|^{\al+m},
\]
which is of the same form as (\ref{normineq1}) with $\al$ replaced by $\al+m$,
and $n$, by $n-m$. Hence, we complete the proof of (\ref{normineq}).

\section*{Acknowledgements}
We are grateful to Alexei Borodin for attracting our attention to this problem.
We also thank Alexander Its and Arno Kuijlaars for useful discussions.  
P. Deift was supported in part by NSF grant \# DMS 0500923.
I. Krasovsky and J. Vasilevska were supported in part by EPSRC grant \# EP/E022928/1.

\end{document}